\documentclass[11pt]{article}
\usepackage{subfigure}
\usepackage{graphics}
\usepackage{amssymb}

\usepackage{float}
\usepackage{url}
\usepackage{graphicx}
\usepackage{amsmath}
\usepackage{amssymb}

\usepackage{makeidx}

\makeindex

\allowdisplaybreaks[2]

\newcommand{\Pomeron}{I\!\!P}

\usepackage{epsf}
\usepackage{epsfig}
\usepackage{rotating}
\usepackage{pstricks}
\usepackage{color}

\def\ds#1{#1\kern-1ex\hbox{/}}
\def\dsh{h\kern-1.2ex /}

\newcommand{\bea}{\begin{eqnarray}}
\newcommand{\eea}{\end{eqnarray}}

\def\beq{\begin{equation}}
\def\eeq{\end{equation}}

\def\beqn{\begin{eqnarray}}
\def\eeqn{\end{eqnarray}}
\def\ba{\begin{eqnarray}}
\def\ea{\end{eqnarray}}

\newcommand{\be}{\begin{equation}}
\newcommand{\beqa}{\begin{eqnarray}}
\newcommand{\eeqa}{\end{eqnarray}}

\newcommand{\ee}{\end{equation}}
\begin{document}

\begin{flushright}
Preprint SMU-HEP-12-17
\end{flushright}
\begin{center}
\vspace{4.cm}

{\bf \Large Massive neutral gauge boson production as a probe of nuclear  
modifications of parton distributions at the LHC}

\vspace{1cm}
{ \bf $^a$Vadim Guzey, $^{b,c}$Marco Guzzi, $^b$Pavel M. Nadolsky, \\ $^d$Mark Strikman, and $^b$Bowen Wang}

{\it 
$^a$ Department of Physics, Hampton University, \\Hampton, VA 23668, USA\\
$^b$Department of Physics, Southern Methodist University, \\Dallas, TX 75275, USA\\
$^c$Deutsches Elektronensynchrotron DESY, \\Notkestrasse 85 D-22607 Hamburg, Germany\\
$^d$Department of Physics, The Pennsylvania State University, \\
State College, PA 16802, USA}

\date{\today}

\begin{abstract}

We analyze the role of nuclear modifications of parton distributions, notably, the
nuclear shadowing and antishadowing corrections, in
production of lepton pairs from decays of neutral $Z$ and $\gamma^{\ast}$ gauge bosons 
in proton-lead and lead-collisions at the LHC.
Using  the Collins-Soper-Sterman resummation formalism that we extended to the case 
of nuclear parton distributions, we observed a direct correlation between the predicted
behavior of the transverse momentum and rapidity distributions of the produced vector bosons
and the pattern of quark and gluon nuclear modifications. 
This makes production of $Z/\gamma^{\ast}$ in $pA$ and $AA$ collisions at the LHC a useful tool
for constraining nuclear PDFs in the small-$x$ shadowing and moderate-$x$ antishadowing regions.

\end{abstract}
\end{center}
\newpage

\section{Introduction}

Nuclear modifications of structure functions and parton distribution functions (PDFs) are a 
firmly established phenomenon extensively studied during last three 
decades~\cite{Frankfurt:1988nt,Arneodo:1992wf,Geesaman:1995yd,Piller:1999wx,Frankfurt:2011cs,Frankfurt:2012qs}.
Presenting the emerging pattern of these modifications in terms
 of the ratio of the nuclear to free 
nucleon structure functions, $R(F_2) \equiv F_{2A}(x)/[A F_{2N}(x)]$, one observes 
the following characteristic trend in different regions of Bjorken $x$:
$R(F_2) < 1$ for $x \leq 0.05-0.1$ (nuclear shadowing),
 $R(F_2) > 1$ for $0.1 \leq x \leq 0.3$ (antishadowing), $R(F_2) < 1$ for $0.3 \leq x \leq 0.8$
(the EMC effect), and $R(F_2) > 1$ for $x > 0.8$ (Fermi motion). 
The four nuclear effects that are responsible for the medium modifications of the PDFs
have different magnitudes and patterns. Indeed, the suppression due to 
nuclear shadowing increases with the atomic number $A$ and could be as large as, {\it e.g.}, 
20\% for $^{40}$Ca; the enhancement due to antishadowing is a few-percent effect that 
does  not reveal a distinctive $A$ dependence; the EMC effect suppression increases with 
$A$ and reaches, {\it e.g.}, 20\% around $x=0.65$ for $^{56}$Fe; 
the enhancement of $R(F_2)$ due to the nucleon Fermi motion inside nuclei for $x> 0.8$ 
increases with $x$, becomes as large as $20-30$\% for $x \sim 0.9$, and leads to 
a nonvanishing nuclear PDF at $x>1$.

In the leading-twist approach,  
this pattern of nuclear modifications of 
$R(F_2)$ translates into a similar picture
of medium modifications of PDFs~\cite{deFlorian:2003qf,Hirai:2007sx,Eskola:2009uj,Schienbein:2009kk,deFlorian:2011fp}. 
However, the magnitude of nuclear modifications of 
individual quark and gluon distributions in nuclei in different regions 
of the light-cone fraction $x$ is rather uncertain. This is especially true 
for the gluon channel, as the indirect extraction of the gluon distribution
from mostly fixed-target experiments on deep inelastic scattering (DIS) 
with nuclear targets leaves the gluon distribution in nuclei 
largely unconstrained.

In this paper we examine the possibility of establishing additional constraints
on nuclear PDFs, primarily in the small-$x$ nuclear shadowing region,
by analyzing the size of nuclear corrections in neutral gauge boson 
production in 
proton-lead and lead-lead reactions at the Large Hadron Collider (LHC). 
In particular, we study the transverse momentum distribution 
of $Z/\gamma^{\ast}$ 
production 
with its subsequent decay into a lepton pair in the reaction 
$A+B\rightarrow(Z/\gamma^{\ast}\rightarrow \ell\bar{\ell})+ X$, 
with $A=p$ or $^{208}\mbox{Pb}$, $B={}^{208}\mbox{Pb}$, and $\ell=e$ or $\mu$,
by using the Collins-Soper-Sterman (CSS) formalism~\cite{Collins:1981uk,Collins:1981va,Collins:1984kg,CollinsBook2} 
implemented in the resummation computer code 
\textsc{ResBos}~\cite{Balazs:1997xd,Landry:2002ix}. 
For this computation, we 
used the nuclear PDFs described in Sec.~\ref{sec:LT_nuclear_shadowing}
and carried out the resummation calculation with the resulting nuclear parton distributions.
The 
choice 
of this channel presents several advantages: it is very clean, 
has a high rate and is very well understood in proton-proton collisions. 
Moreover, the $Q_{T}$ distribution of heavy vector bosons 
can be easily reconstructed experimentally and, thus, 
can be measured very precisely. Access to a wide range of boson virtualities 
provides a powerful test of the leading-twist factorization and QCD evolution. 

The extension of the CSS resummation formalism to nuclear scattering is described 
in Sec.~\ref{sec:QTResum}. The treatment 
of nuclear corrections in the leading-twist approach is reviewed in
Sec.~\ref{sec:LT_nuclear_shadowing}, while
Sec.~\ref{sec:NumericalPredictions} presents numerical predictions
for nuclear corrections in $Z/\gamma^*$ production in $pA$ and $AA$
scattering at the LHC, 
in two distinct intervals of the Drell-Yan pair mass, 
$70 < Q < 110$ GeV and $5 < Q < 20$ GeV. We examine nuclear corrections 
to the Drell-Yan rapidity and transverse momentum distributions and discuss 
dependence on the nonperturbative $k_T$ smearing corrections.
Finally, Sec.~\ref{sec:conclusions} contains our conclusions.

\section{Application of $Q_T$ resummation to nuclear
  scattering\label{sec:QTResum}}

\subsection{$pp$ collisions}

In the standard framework \cite{Collins:1981uk,Collins:1981va,Collins:1984kg,CollinsBook2} for transverse momentum
resummation in nucleon-nucleon collisions,
the differential cross section for vector boson production reads
\begin{eqnarray}
\frac{d\sigma}{dQ^{2}dydq_{T}^{2}} & = & \int\frac{d^{2}\vec{b}}{(2\pi)^{2}}e^{i\vec{q}_{T}\cdot\vec{b}}\widetilde{W}(b,Q,y)=
\nonumber \\
 & = & \int_{0}^{\infty}\frac{bdb}{2\pi}J_{0}(q_{T}b)\widetilde{W}^{pert}(b_{*},Q,y)\widetilde{W}^{NP}(b,Q,y)+Y(Q,Q_{T},y),
\label{Res}
\end{eqnarray}
where $Q,$ $Q_{T}$, and $y$ are the invariant mass, transverse
momentum, and rapidity of the vector boson, respectively. The cross
section consists of two terms, a Fourier-Bessel integral over the
transverse position ($b$) that evaluates the resummed logarithmic
contributions in the limit $Q_{T}^{2}\ll Q^{2}$, and the finite piece
$Y(Q,Q_{T},y$) that dominates at $Q_{T}^{2}\approx Q^{2}$. 
The resummed form factor $\widetilde{W}(b,Q,y)$
receives leading-power and power-suppressed contributions, and those
are incorporated into the resummation integral of Eq.~(\ref{Res})
as $\widetilde{W}^{pert}$ and $\widetilde{W}^{NP}$, respectively.

The form factor $\widetilde{W}^{pert}$ and non-singular contribution
$Y$ (defined as the difference between the fixed-order calculation
and the approximation for the perturbative part $\widetilde{W}^{pert}$
to the same order in $\alpha_{s}$) are computed in perturbative QCD.
The explicit expression for $\widetilde{W}^{pert}$ is 
\begin{eqnarray}
\widetilde{W}^{pert} & = & \sum_{j=u,d,s...}\sigma_{j}^{(0)}\, e^{-{\cal S}(b,Q,C_{1},C_{2})}
\nonumber \\
 & \times & \sum_{a=g,q}\left[{\cal C}_{ja}\otimes f_{a/h_{1}}\right]\left(x_{1},\frac{C_{1}}{C_{2}},
\frac{C_{3}}{b}\right)\sum_{b=g,q}\left[{\cal C}_{\bar{j}b}\otimes f_{b/h_{2}}\right]
\left(x_{2},\frac{C_{1}}{C_{2}},\frac{C_{3}}{b}\right)
\label{Wpert}
\end{eqnarray}
in terms of the Sudakov exponential 
\be
e^{-{{\cal S}}(b,Q,C_{1},C_{2})}=\exp\left[-\int_{C_{1}^{2}/b^{2}}^{C_{2}^{2}Q^{2}}\frac{d\bar{\mu}^{2}}{\bar{\mu}^{2}}{\cal A}(\bar{\mu};C_{1})\,\ln\left(\frac{C_{2}^{2}Q^{2}}{\bar{\mu}^{2}}\right)+{\cal B}(\bar{\mu};C_{1},C_{2})\right]
\label{Sudakov}
\ee
and convolutions $\left[{\cal C}_{j/a}\otimes f_{a/h}\right]$ of
Wilson coefficient functions ${\cal C}_{j/a}(z,\, C_{1}/C_{2},\,\mu_{F}=C_{3}/b)$
with distributions $f_{a/h}(z,\mu_{F})$ of partons
$a$ inside the initial-state hadron $h$. The normalization factor
is 
\begin{eqnarray}
\sigma_{j}^{(0)} & \equiv & \frac{M_Z^2\,G_F^2}{12 N_c S}\frac{Q^2}{(Q^2-M_Z^2)^2+M_Z^2
  \Gamma_Z^2}\left[\left(1-4\left|e_{j}\right|\sin^{2}\theta_{w}\right)^{2}+1\right]\nonumber\\
&\times & \left[\left(1/2 - \sin^{2}\theta_{w}\right)^{2}+\sin^4\theta_w \right]\label{sigma0Z}
\end{eqnarray}
in $pp\rightarrow (Z^{0}\rightarrow e^+e^-)X$,
and 
\begin{equation}
\sigma_{j}^{(0)}\equiv\frac{1}{S}\frac{4\pi^{2}\alpha_{EM}^{2}}{3N_{c}Q^{2}}e_{j}^{2}\label{sigma0DY}
\end{equation}
in $pp\rightarrow(\gamma^{*}\rightarrow e^+ e^-)X.$
Here $M_Z$ is the $Z$ boson mass, $G_{F}$ is the Fermi constant, $\theta_{w}$ is the weak angle,
$\sqrt{S}$ is the collision energy, $N_{c}=3$ is the number of colors, $\alpha_{EM}$ is the fine structure constant,
 and $e_{j}$=2/3 or -1/3 is the
fractional quark charge. The scale dependence of
$\widetilde{W}^{pert}$ is ascribed to three scale parameters $C_{1}=b\bar{\mu}$,
$C_{2}=\bar{\mu}/Q$, where $\bar{\mu}$ is the integration variable
in the Sudakov exponent, and $C_{3}=\mu_{F}b,$ where $\mu_{F}$ is
the factorization scale in the convolutions $\left[{\cal C}_{j/a}\otimes f_{a/h}\right]$
with the PDFs.

The function $\widetilde{W}^{NP}(b,Q)$ in Eq.~(\ref{Res}) provides
a model for power-suppressed terms in the resummed form factor that
play an important role at transverse positions of order $1$ GeV$^{-1}$
or above. We evaluate the perturbative part $\widetilde{W}^{pert}$ as a function
of $b_{*}=b/\sqrt{1+b^{2}/b_{max}^{2}}$ and combine with the nonperturbative
function $\widetilde{W}^{NP}(b,Q)$. The 
$b_{*}$ variable \cite{Collins:1981va,Collins:1984kg} realizes a smooth 
transition from the leading-power terms that
dominate at small $b$ to nonperturbative dynamics that is important 
at large $b$. We implement the $b_*$ convention in its extended version
\cite{Konychev:2005iy} that can simulate various trends in the 
behavior of $\widetilde{W}(b,Q)$ at $b>1 \mbox{ GeV}^{-1}$ suggested by the
nonperturbative models
\cite{Guffanti:2000ep,Qiu:2000hf,Tafat:2001in,Kulesza:2002rh,Becher:2011xn,Echevarria:2012pw},
while at the same time retaining the exact perturbative prediction for 
$\widetilde{W}(b,Q)$ at $b< 1 \mbox{ GeV}^{-1}$. 

Parameters of $\widetilde{W}^{NP}(b,Q)$ are constrained by a fit to
the experimental data. To reproduce the observed $Q_T$ distributions
in Drell-Yan process, it
suffices to approximate $\widetilde{W}^{NP}(b,Q)$ by a Gaussian smearing
factor $\exp\left[-a(Q,x_{1},x_{2})b^{2}\right]$, where the $b^{2}$
dependence is associated with the lowest power-suppressed contribution
\cite{Korchemsky:1994is}, and $a(Q,x_{1},x_{2})$ is found from the 
fit \cite{Ladinsky:1993zn,Landry:2002ix}. A recent parametrization 
for $a(Q)$ from \cite{Konychev:2005iy}
describes well both the low-$Q$ pair and $Z$
production. Recently, constraints on $a(Q)$ at $Q\approx M_{Z}$
were updated \cite{Guzzi:2012jc,GNW} using data on angular distributions
$d\sigma/d\phi_{\eta}^{*}$ in $Z$ boson production at the Tevatron
Run-2~\cite{Abazov:2010mk}. This updated nonperturbative
parametrization will be used in our simulations, as it provides 
excellent description of $Z$ boson $Q_T$ distributions in $pp$ collisions at the
LHC.

\subsection{$pA$ and $AA$ collisions}
The CSS resummation formalism can be adapted for describing nuclear
collisions using the leading-twist approach discussed in the next
section. Since the hard scattering contributions are identical in
proton and nuclear collisions in the leading-twist approximation,
it suffices to replace the proton PDFs $f_{a/h}(x,\mu_{F})$ in all
parts of the resummed cross section (\ref{Res}) by nuclear PDFs
$f_{a/A}(x,\mu_{F})$ for each initial-state nucleus. The smearing
function $\widetilde{W}^{NP}(b,Q)$ may also depend on the type of the
nuclear projectile, for example, due to nuclear broadening 
effects \cite{Kang:2008us},
and can be modified by adjusting the $a$ parameter.

Following this approach, the resummed predictions can be obtained for any
set of nuclear PDFs 
that is available \cite{deFlorian:2003qf,Hirai:2007sx,Eskola:2009uj,Schienbein:2009kk,deFlorian:2011fp}.
In the next Section, we describe the FGS nuclear PDFs that will be
used in our numerical simulations.

\section{Leading-twist nuclear shadowing in nuclear parton distributions}
\label{sec:LT_nuclear_shadowing}
\iffalse
%%%%%%%%% insert author names addresses here, please do not change format

\hspace{\parindent}\parbox{\textwidth}{\slshape 
  Vadim Guzey$\,^1$, Mark Strikman$\,^2$ \\[1ex]
%
$^1$\ Theory Center, Jefferson Lab, Newport News, VA, USA \\
$^2$\ Physics Department, Pennsylvania State University, State College , PA,
  USA
}

% put the author names here again, for the list of authors at the end of
% the volume
\index{Guzey, Vadim}
\index{Strikman, Mark}
\fi

%\vspace{2\baselineskip}
\subsection{Theory roundup}
Nuclear shadowing in hadron-nucleus and real photon-nucleus scattering
is a firmly established experimental phenomenon. 
At high energies, the scattering cross
section on a nuclear target is smaller than the sum of the cross sections
on the individual nucleons comprising the nuclear target.
In the nucleus rest frame, nuclear shadowing is explained as being an example of quantum
mechanical destructive interference between the scattering amplitudes corresponding to
the interaction with one, two, three, etc., nucleons~\cite{Glauber:1955qq}.
The modern theory of nuclear shadowing is based on the connection between 
nuclear shadowing and 
diffraction~\cite{Gribov:1968jf}, which can be understood~\cite{Frankfurt:1998ym} 
as a manifestation of unitarity reflected in the Abramovsky-Gribov-Kancheli (AGK) 
cutting rules~\cite{Abramovsky:1973fm}.
The accuracy of the resulting theory of nuclear shadowing is rather high, with corrections
at the level of no more than a few percent.

Turning to hard processes with nuclei, it was observed that
the connection between shadowing and diffraction is also valid
for any processes initiated by a hard probe and, in  particular,  for 
deep inelastic scattering (DIS) with nuclei.
Combining this connection with QCD factorization theorems for inclusive and 
diffractive DIS, nuclear parton distribution functions (PDFs) for small 
$x$ ($x < 0.05)$ can be calculated. 
The approach based on this connection is called the leading-twist theory
of nuclear shadowing~\cite{Frankfurt:2011cs,Frankfurt:1998ym,Frankfurt:2003zd,Guzey:2009jr}
%vg
and its predictions for nuclear PDFs will be used in the numerical analysis in this paper.

\begin{figure}
\begin{center}
\includegraphics[width=0.8\textwidth]{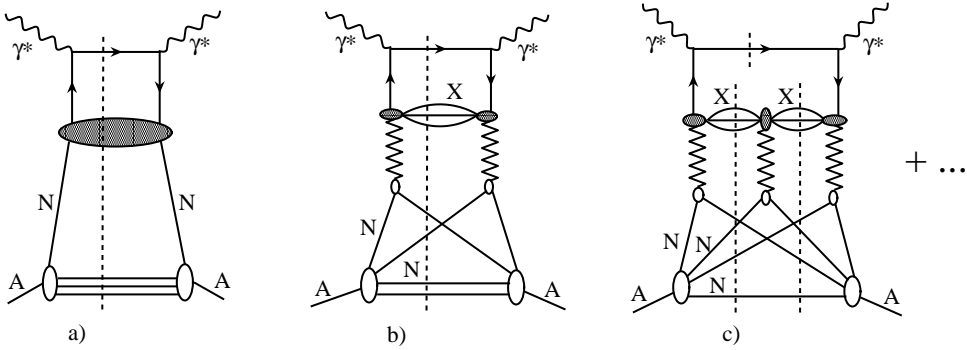}
\end{center}
\caption
{\label{fig:gfs10_Master1_quarks} Multiple scattering series for nuclear quark PDFs. 
Graphs $a$, $b$, and $c$ correspond to the interaction with one, two, and three
nucleons, respectively. 
Graph $a$ gives the impulse approximation;
graphs $b$ and $c$ contribute to the shadowing correction;
the interaction with more than three nucleons is not shown but included in the final answer. 
}
\end{figure}

The modifications  of nuclear parton distributions due to nuclear shadowing 
are
given by a multiple scattering series, where each term corresponds to the interaction with
one, two, three, etc.~nucleons. 
In the graphical form, 
an example of such a series for the quark distributions in nuclei is
presented in Fig.~\ref{fig:gfs10_Master1_quarks}. 
The corresponding expression for the nuclear PDF of flavor $j$ at a certain initial scale 
$Q_0^2$ reads~\cite{Guzey:2009jr,Frankfurt:2011cs}:
%\begin{eqnarray}
\bea
&&xf_{j/A}(x,Q_0^2)=Axf_{j/N}(x,Q_0^2) \nonumber\\
&-&8 \pi A (A-1)\, \Re e \frac{(1-i \eta)^2}{1+\eta^2}
\int^{0.1}_{x} d x_{\Pomeron} \beta f_j^{D(4)}(\beta,Q_0^2,x_{\Pomeron},t_{\rm min})\int d^2 b \int^{\infty}_{-\infty}d z_1 \int^{\infty}_{z_1}d z_2
 \nonumber\\
&\times& \rho_A(\vec{b},z_1) \rho_A(\vec{b},z_2) e^{i (z_1-z_2) x_{\Pomeron} m_n}
 e^{-\frac{A}{2} (1-i\eta) \sigma_{\rm soft}^j(x,Q_0^2) \int_{z_1}^{z_2} dz^{\prime} \rho_A(\vec{b},z^{\prime})} \,,
\label{eq:fgs10_eq2}
%\end{eqnarray}
\eea
where $f_j^{D(4)}$ is the diffractive parton distribution of the nucleon; 
$\rho_A$ is the nuclear matter density; $\eta$ is the ratio of the real to imaginary
parts of the elementary diffractive amplitude, 
$\eta =\Re e A^{\rm diff}/ \Im m A^{\rm diff} \approx 0.17$. 
The diffractive PDF $f_j^{D(4)}$ depends on two light-cone fractions
$x_{\Pomeron}=(M_X^2+Q^2)/(W^2+Q^2)$ and $\beta=x/x_{\Pomeron}$, and the invariant
momentum transfer $t$, where $W$ is the invariant virtual photon-nucleon energy, 
$W^2=(q+p)^2$, and $M_X^2$ is the invariant mass squared of the produced intermediate
diffractive state, denoted as ''X'' in Fig.~\ref{fig:gfs10_Master1_quarks}.
The longitudinal (collinear with the direction of the photon momentum) 
coordinates $z_1$ and $z_2$ and the transverse coordinate (impact parameter) $\vec{b}$ 
refer to the two interacting nucleons; $m_n$ is the nucleon mass.
The $t$ dependence of $f_j^{D(4)}$ can be safely neglected as compared to the strong 
fall-off of the nuclear form-factor for $A> 4$. As a result, $f_j^{D(4)}$ enters 
Eq.~(\ref{eq:fgs10_eq2}) at $t_{\rm min}\approx -x ^2 m_n^2(1+M_X^2/Q^2)^2$, and all 
nucleons enter with the same impact parameter $\vec{b}$.

In Fig.~\ref{fig:gfs10_Master1_quarks} and in Eq.~(\ref{eq:fgs10_eq2}), 
modification of nuclear PDFs due to  the interaction with two nucleons is calculated 
in a model-independent way through the nucleon diffractive PDFs, similarly to the case of 
hadron-deuteron scattering in the Gribov-Glauber theory of  shadowing in soft processes.
The contribution 
to nuclear shadowing from the interactions with $n\ge 3$ nucleons 
requires
additional model-dependent considerations,
since the interaction of a hard probe (virtual photon) with $n \ge 3$ nucleons is sensitive 
to fine details of the diffractive dynamics. Noticing that the analysis of diffraction in DIS 
implies that hadronic fluctuations of the virtual photon have predominantly large sizes, 
one can reliably parameterize  the strength of the 
interaction with $n\ge 3$ nucleons by a single effective hadron-like cross section $\sigma_{\rm soft}^j$.
It was argued in Ref.~\cite{Frankfurt:2011cs} that the magnitude of
$\sigma_{\rm soft}^j$ can be reliably estimated using either 
the color dipole model or the model for hadronic fluctuations of the pion.
As a result, using these two approaches one effectively obtains 
the lower and upper boundaries on $\sigma_{\rm soft}^j$, which in turn
correspond to the upper (FGS10\_H) and lower (FGS10\_L) boundaries on the 
predicted nuclear shadowing.

Note that Eq.~(\ref{eq:fgs10_eq2}) defines $xf_{j/A}(x,Q^2_0)$ only for the shadowing region of
$x \leq 0.1$ and at $Q_0=2$~GeV~~\cite{Frankfurt:2011cs}. In the $0.03 \leq x \leq 0.2$ interval, 
there was introduced an additional effect of nuclear antishadowing (enhancement)
in the gluon channel and modeled by requiring conservation of the momentum
sum rule. Finally, while it was assumed that in the gluon and sea quark channels
$xf_{j/A}(x,Q^2_0)=Axf_{j/N}(x,Q^2_0)$ for $x > 0.2$, the valence quarks are 
subject to all four types of nuclear modifications briefly discussed in 
the introduction. As a consequence of the DGLAP $Q$ evolution from $Q_0$ to
$Q > Q_0$, the large-$x$ nuclear effects, notably, the EMC effect, 
start also affecting the gluon and sea quark nuclear PDFs, see Figs. 2 and 3.

\subsection{Practical implementation}
We write a nuclear PDF as
\ba
f_{j/A}(x,Q^{2})\equiv R_{j}(x,Q^{2})\cdot(A f_{j/N}(x,Q^{2})) \,,
\label{pdfratio}
\ea
where $R_{j}(x,Q^{2})$ is the ratio given by the leading-twist theory of nuclear corrections;
$f_{j/N}$ is the free nucleon PDF.
For $j=u,\bar{u},d$, and $\bar{d}$ we have 
\ba
Af_{j/N}(x,Q^{2}) \equiv Zf_{j/p}(x,Q^{2}) + (A-Z) f_{j/n}(x,Q^{2}) \,,
\ea
where $A$ is the atomic number, $Z$ is the number of protons, $A-Z$ 
is the number of neutrons. The subscripts $p$ and $n$ indicate a proton and a neutron, respectively.
The PDFs for the neutron for $u$ and $d$ (anti-)quarks 
are obtained via charge symmetry, $f_{u/n}(x,Q^2) = f_{d/p}(x,Q^2)$, etc.
For heavier quarks and gluons, we have
\ba
f_{j/N}(x,Q^{2})=f_{j/p}(x,Q^{2}) \,,\ 
\hspace{1cm} j=s,\bar{s},c,\bar{c},b,\bar{b},g.
\ea

\begin{figure}[p]
\begin{center}
%\hspace{-1cm}
\subfigure[]{
\includegraphics[width=0.45\textwidth]{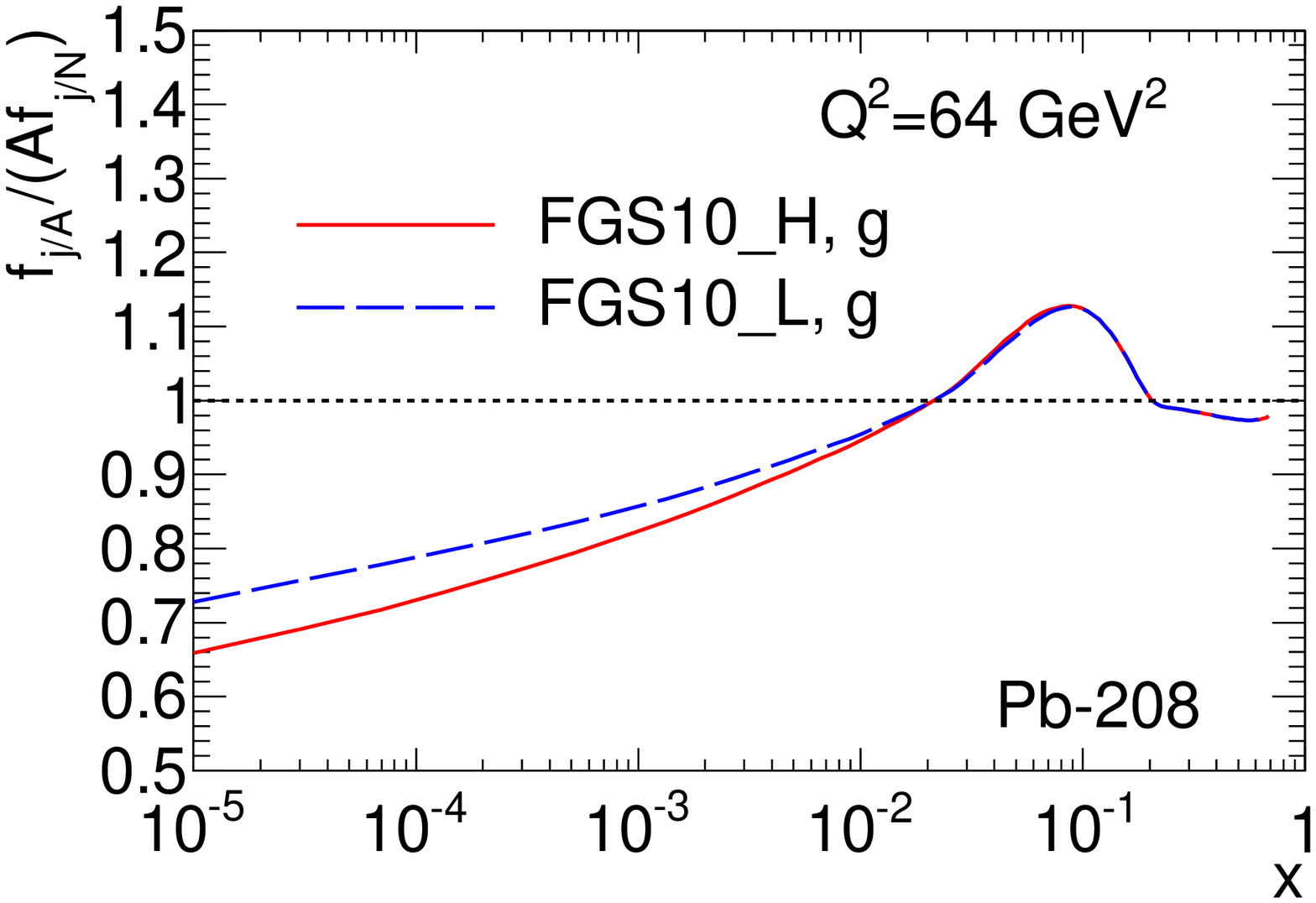}}
\subfigure[]{
\includegraphics[width=0.45\textwidth]{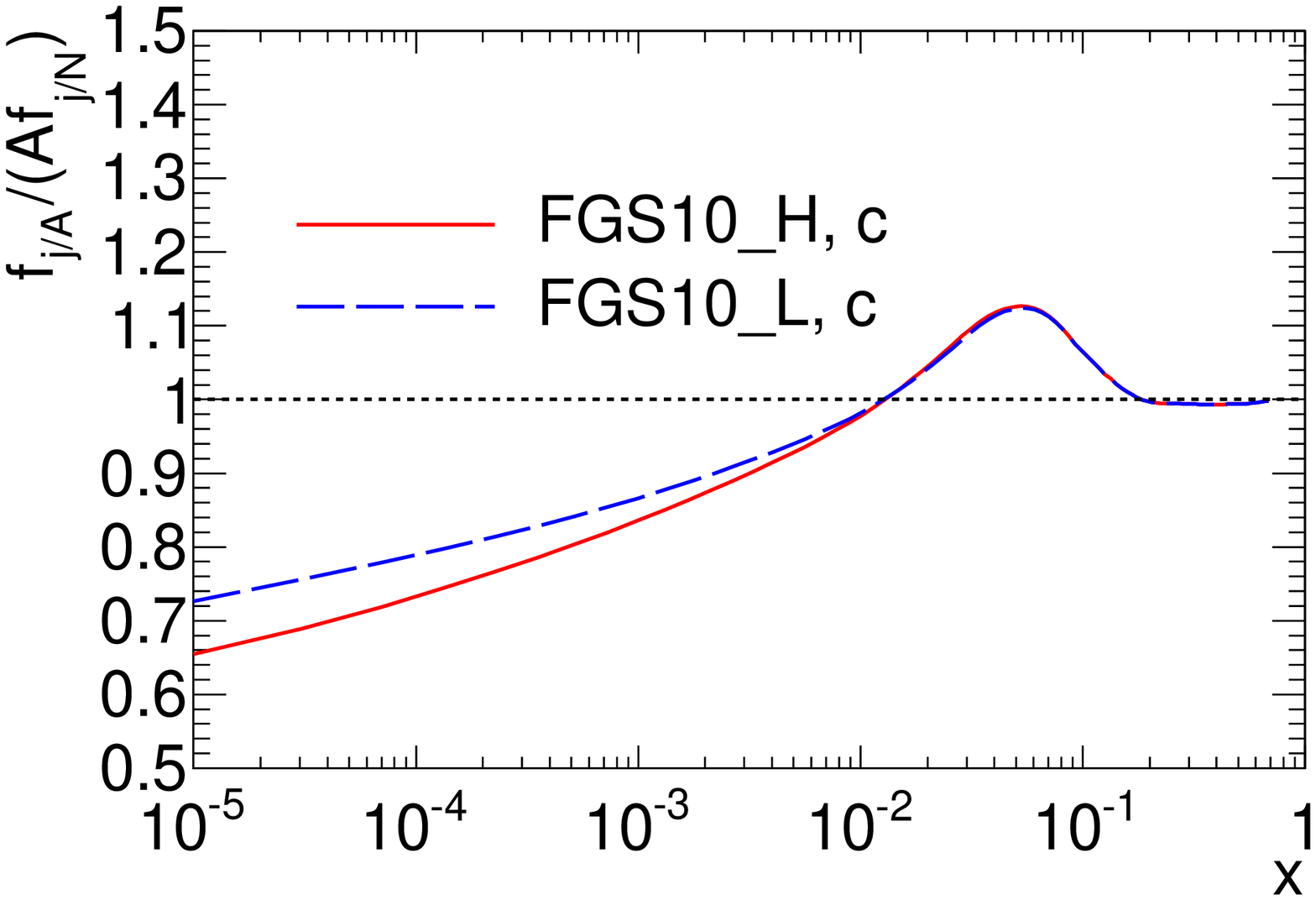}}
\subfigure[]{
\includegraphics[width=0.45\textwidth]{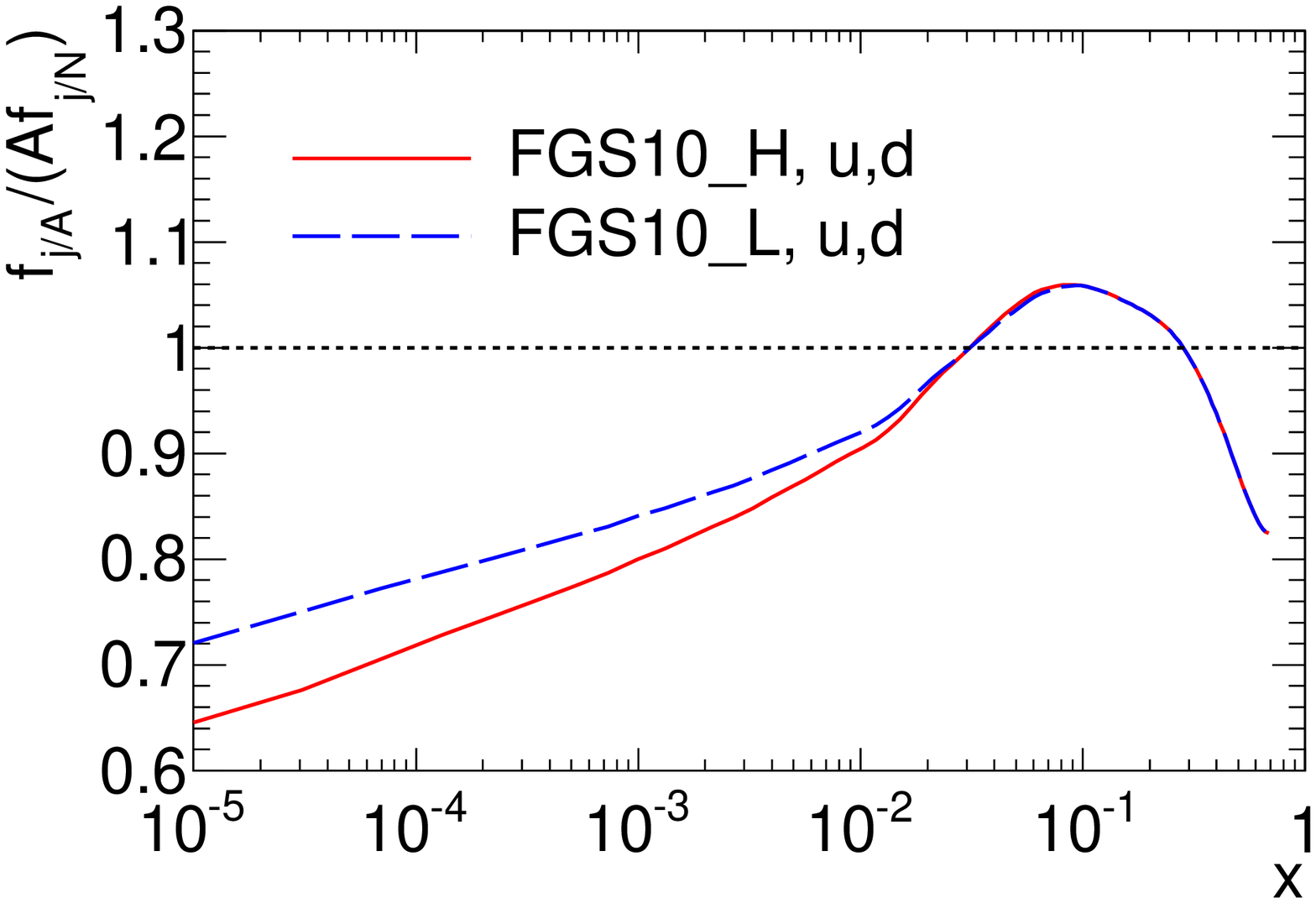}}
\subfigure[]{
\includegraphics[width=0.45\textwidth]{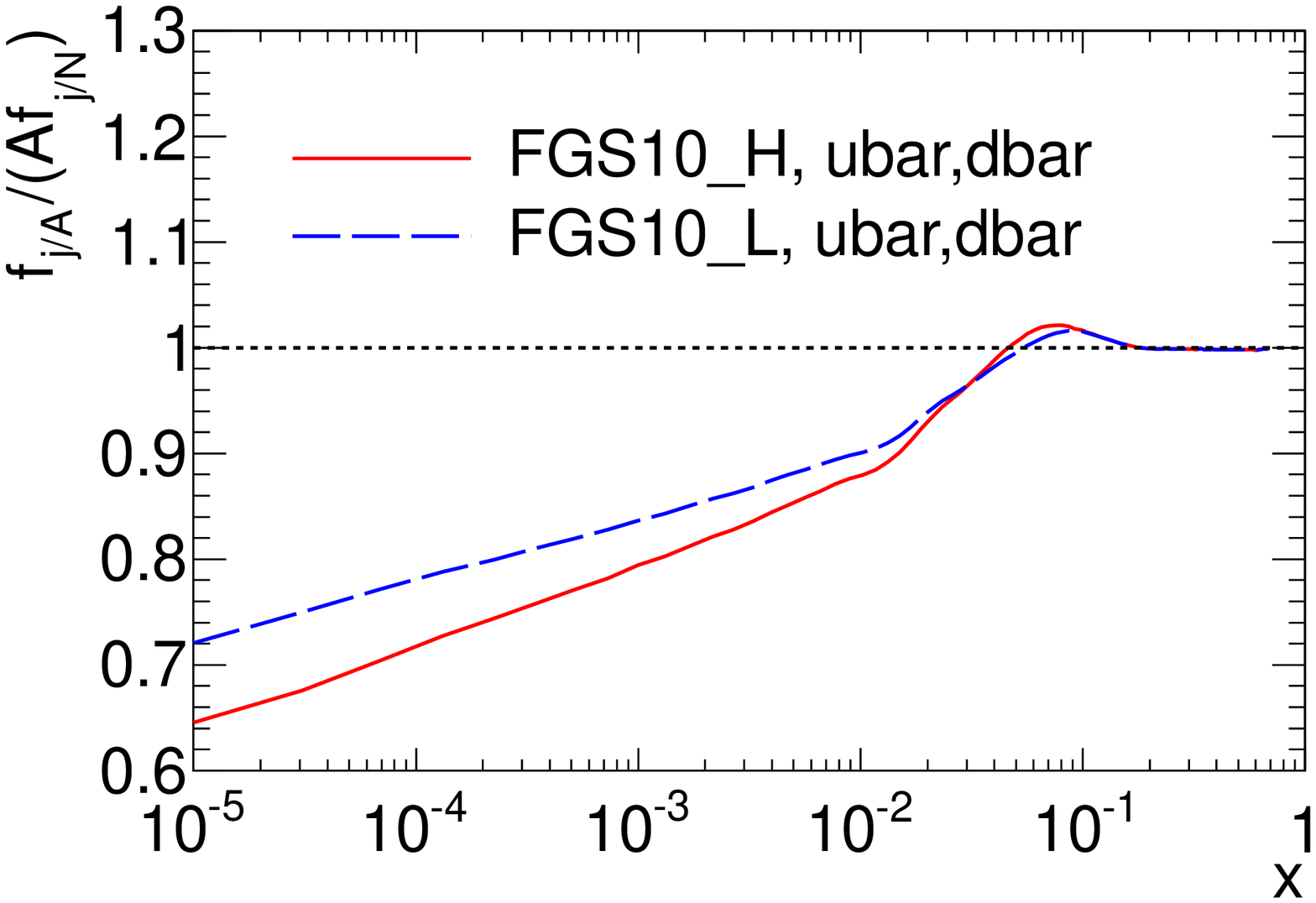}}
\caption{Predictions of the leading-twist theory of nuclear corrections 
for $f_{j/A}(x,Q)/[A f_{j/N}(x,Q^2)]$ for different partons in $^{208}_{82}$Pb, with $Q=8$ GeV, and for 
the indicated parton flavors.
The two sets of curves correspond to the two 
scenarios of nuclear shadowing (see the text).}
\label{pdfratio8}
\end{center}
\end{figure}

\begin{figure}[p]
\begin{center}
%\hspace{-1cm}
\subfigure[]{
\includegraphics[width=0.45\textwidth]{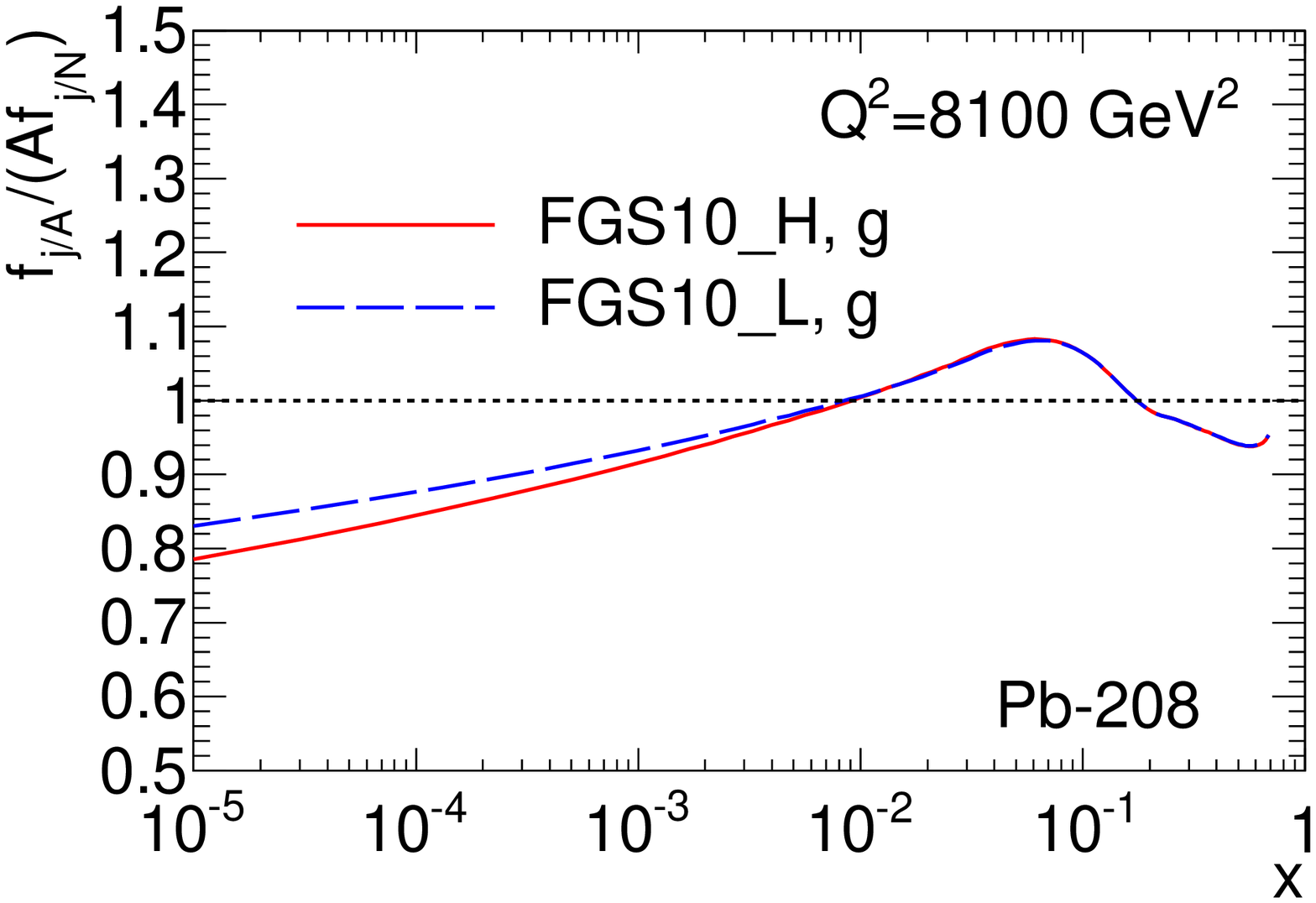}}
\subfigure[]{
\includegraphics[width=0.45\textwidth]{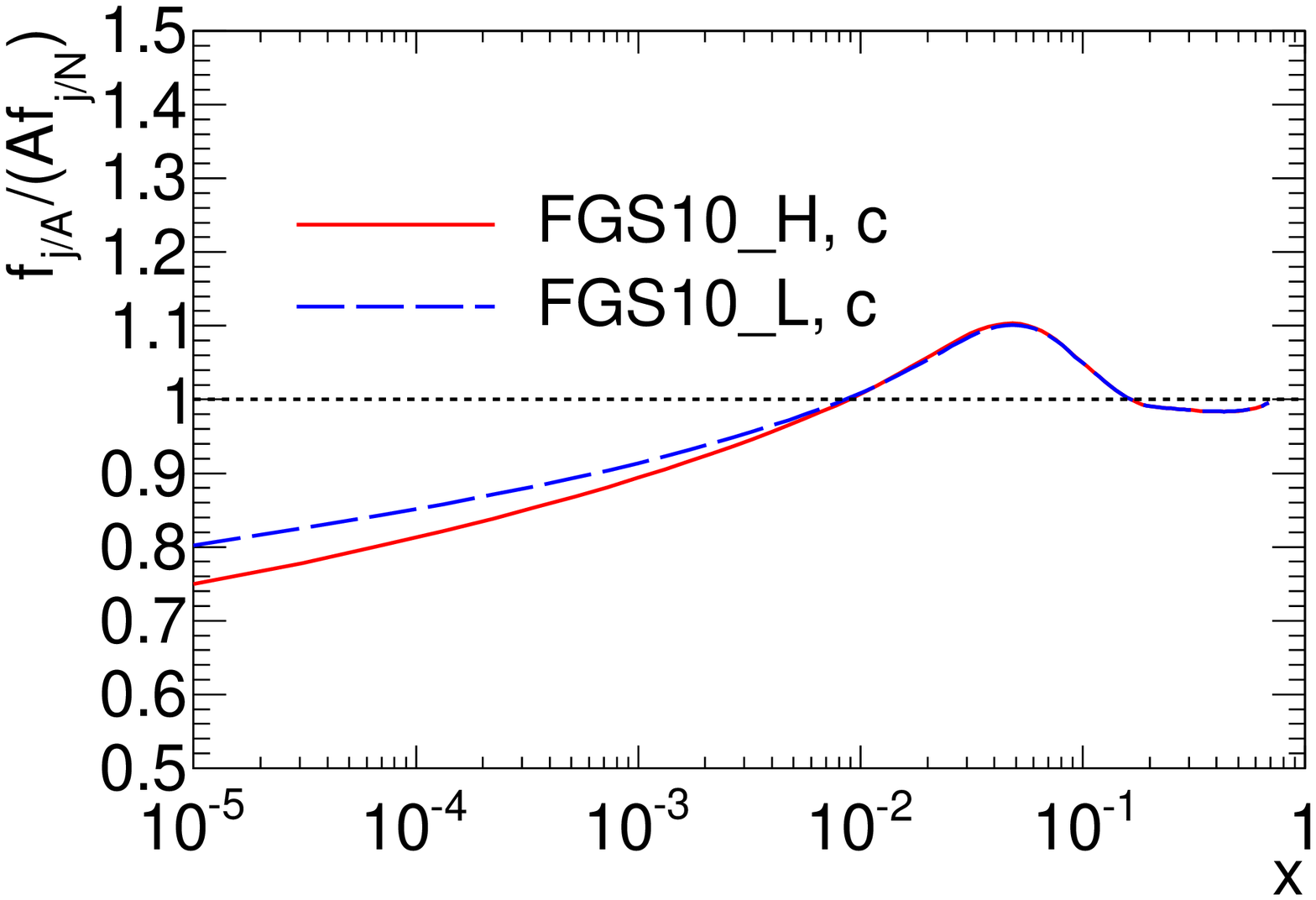}}
\subfigure[]{
\includegraphics[width=0.45\textwidth]{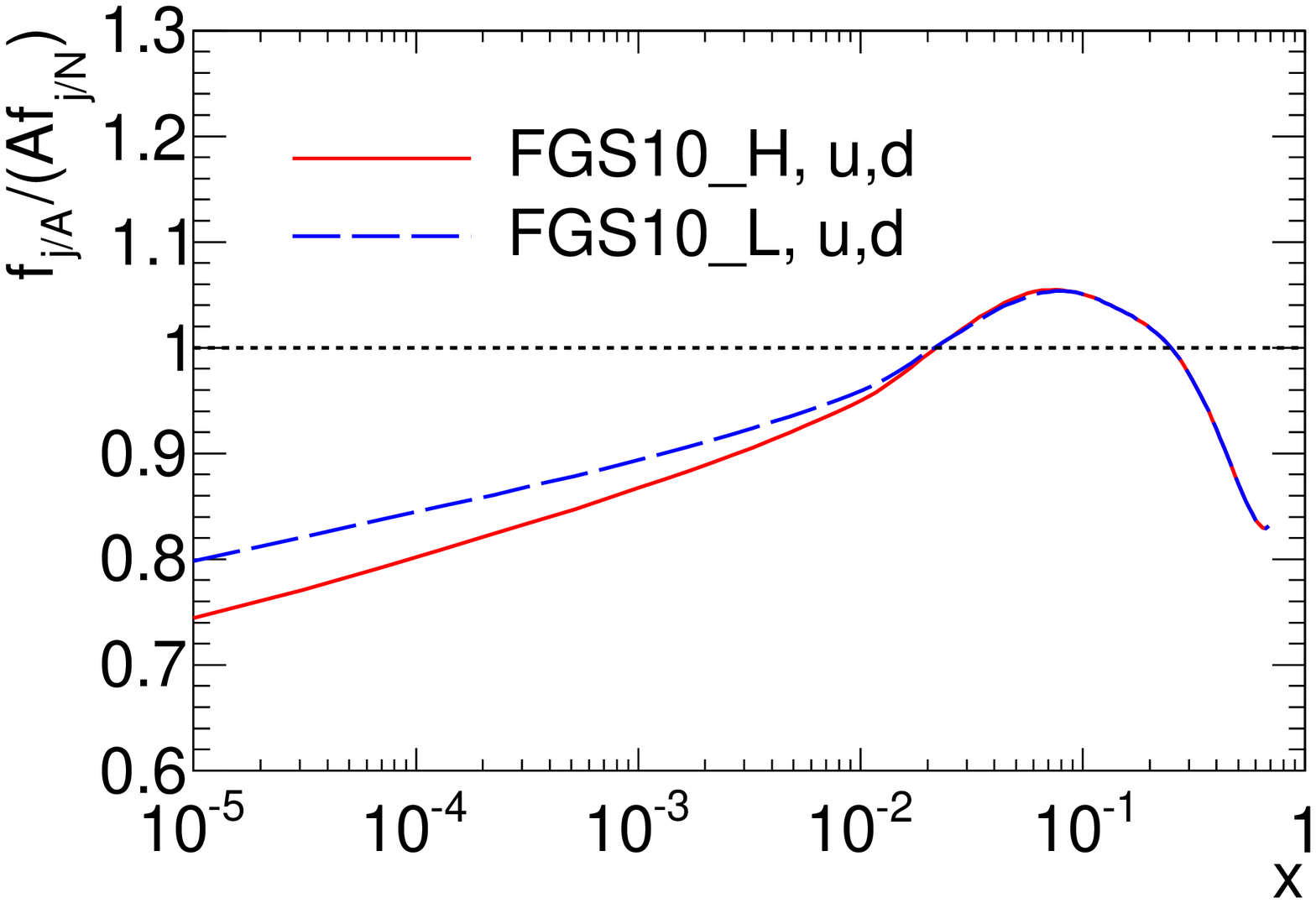}}
\subfigure[]{
\includegraphics[width=0.45\textwidth]{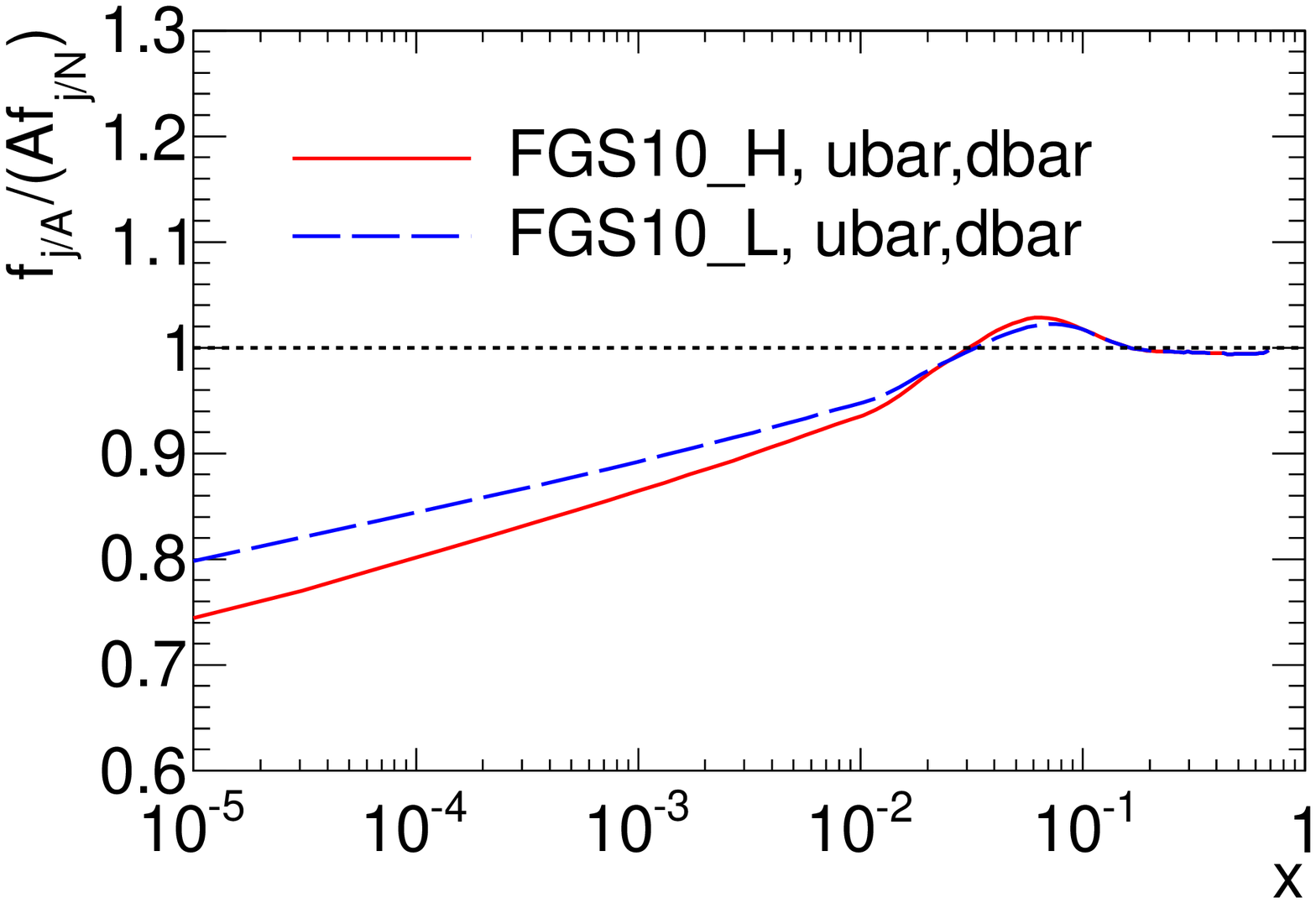}}
\caption{Same as Fig.~\ref{pdfratio8}, for $Q=90$ GeV.}
\label{pdfratio90}
\end{center}
\end{figure}

Examples of resulting predictions of the leading-twist theory of nuclear corrections
are presented in Figs.~\ref{pdfratio8} and \ref{pdfratio90} , where 
the ratios of the next-to-leading order (NLO) nuclear 
($^{208}$Pb) to nucleon PDFs,
 $f_{j/A}(x,Q^2)/[A f_{j/N}(x,Q^2)]$,
is plotted as functions of Bjorken $x$ at $Q=8$ and  $90$ GeV, 
the representative virtualities of the $\gamma^*/Z$ boson encountered in 
our numerical calculation.
The two sets of curves 
(labeled FGS10\_H and FGS10\_L) correspond to the 
two scenarios for $\sigma_{\rm soft}^j$ (nuclear shadowing) that we have mentioned above.
At $Q$ values shown in the figures, 
one predicts
significant medium modifications of nuclear PDFs 
for each parton flavor:
large shadowing (up to 30-40\%) 
for 
$x \lesssim 10^{-2}$,
 antishadowing at $0.05 \lesssim x \lesssim 0.15$, and 
the suppression originating from the EMC effect for
$x \gtrsim 0.3$ values.
The spread between the solid and dotted curves in Figs. ~\ref{pdfratio8} and \ref{pdfratio90}
is the theoretical uncertainty of the present approach to nuclear PDFs; it is less
than  10\% for $A\sim 200$ and is much smaller for light nuclei.

\subsection{Experimental tests in the Drell-Yan process  \label{sec:exptests}}
Next, we wish to outline how the $x$ dependence of the nuclear
correction can be investigated experimentally 
by taking advantage of the basic kinematics of the Drell-Yan process. 
The hadronic cross section $\sigma$ for Drell-Yan pair
production involves convolutions over longitudinal momentum fractions 
$\xi_1, \xi_2$ of the hard scattering cross section
$\widehat \sigma$ with the PDFs, 
\ba
\frac{d\sigma}{dQ^2\,dy\,dQ_T^2}&=&\sum_{a,b}\int_0^1 d\xi_1 \int_0^1 d\xi_2\frac{d\hat{\sigma}}{dQ^2dydQ_T^2}f_{a/A}(\xi_1)
f_{b/B}(\xi_2) 
\nonumber\\
&\equiv&\int_{\xi_{1,min}}^1 d\xi_1 \int_{\xi_{2,min}}^1 d\xi_2\, h(\xi_1,\xi_2)
~\delta\left[\left(\frac{\xi_1}{x_1}-1\right) \left(\frac{\xi_2}{x_2} - 1\right) - \frac{Q_T^2}{M_T^2}\right].
\label{goodlim}
\ea
In the second line, the Dirac delta function is separated
from all other terms, indicated collectively by $h(\xi_1,\xi_2)$.
It arises due to energy-momentum conservation  
and constrains the integration variables $\xi_1$ and
$\xi_2$ to lie on a hyperbolic integration contour satisfying
$K_1 K_2 =Q_T^2/M_T^2$, where $K_{1,2}=\xi_{1,2}/x_{1,2}-1$, 
$x_{1,2}=(M_T/\sqrt{S})\exp(\pm y)$, and $M_T=\sqrt{Q^2 + Q_T^2}$. 

Although $\xi_1$ and $\xi_2$ are integrated
over an extended range, a large part of the cross section is
contributed from around the point $K_1=K_2 = Q_T/M_T$, where both
$\xi_1$ and $\xi_2$ are small enough, and the corresponding PDFs are
large. At this point, 
\begin{eqnarray}
&& \xi_1 \approx x_{1c} \equiv \tau_{eff} e^{y}, \nonumber \\
&& \xi_2 \approx x_{2c} \equiv \tau_{eff} e^{-y},
\label{typx}
\end{eqnarray}
where 
\be
\tau_{eff} \equiv \frac{M_T + Q_T}{\sqrt{S}}.
\ee
Far away from this point, one of the PDFs is suppressed as 
$\xi \rightarrow 1$, and the cross section is reduced.

\begin{figure}[ht]
\begin{center}
\includegraphics[width=0.7\textwidth]{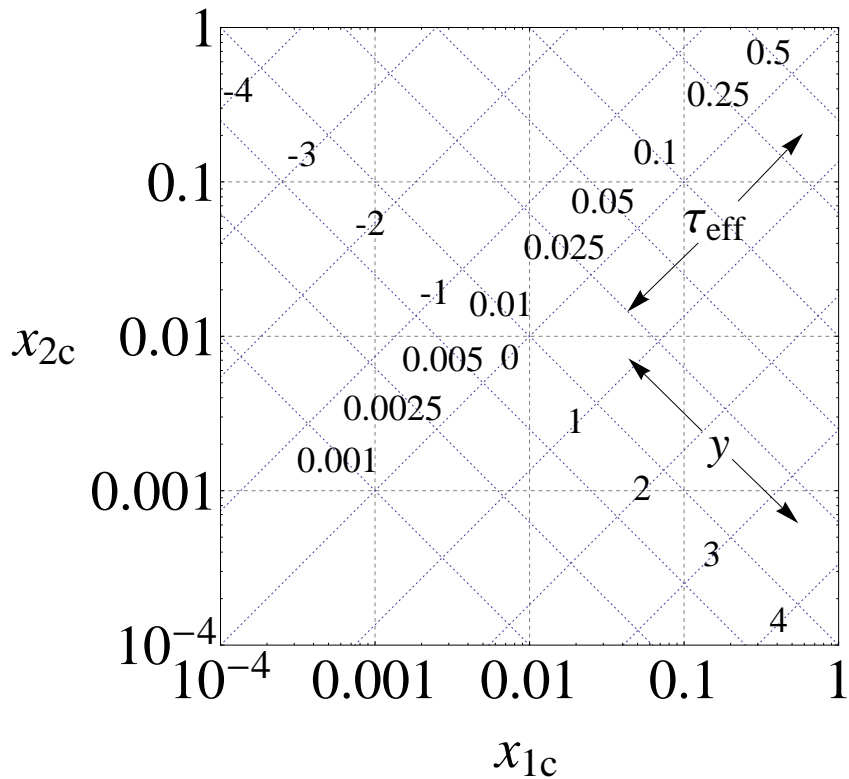}
\caption{Representative momentum fractions $x_{1c}$ and $x_{2c}$ in
  initial-state particles 1 and 2 as functions of $y$ and $\tau_{eff}$.
\label{kine}
}
\end{center}
\end{figure}

The pair $\{x_{1c}, x_{2c}\}$ can thus characterize typical values of
the momentum fractions in the initial-state hadrons 1 and 2 
in most of contributing events for the given boson's 4-momentum. 
They can be evaluated from $\tau_{eff}$ and $y$ 
(or, equivalently, $\sqrt{S}$, $Q$, $Q_T$, and $y$) 
with the help of Eq.~(\ref{typx}).
A contour plot for conversion of $\{\tau_{eff}, y\}$ 
into $\{x_{1c}, x_{2c}\}$ is presented in
Fig.~\ref{kine}. Here, the contours indicate constant values of 
$\tau_{eff}$ and $y$; the corresponding $x_{1c}$ and $x_{2c}$ values 
can be read off the horizontal and vertical axis. 

For the following discussion, we also include Table~\ref{tab:taueff} 
listing the ranges of $\tau_{eff}$ for the combinations of $\sqrt{S}$, $Q$ and $Q_T$ values
typical for our figures. 

\begin{table}
\begin{centering}
\begin{tabular}{|c|c|c|c|}
\hline 
$\sqrt{S},$ TeV & $Q$, GeV & $Q_{T}$, GeV & $\tau_{eff}$\tabularnewline
\hline
2.76 & 8 & 0-100 & 0.0029-0.073\tabularnewline
\cline{2-4} 
 & 90 & 0-200 & 0.033-0.15\tabularnewline
\hline 
4.4 & 8 & 0-100 & 0.0018-0.046\tabularnewline
\cline{2-4} 
 & 90 & 0-200 & 0.020-0.095\tabularnewline
\hline 
5.52 & 8 & 0-100 & 0.0014-0.036\tabularnewline
\cline{2-4} 
 & 90 & 0-200 & 0.016-0.08\tabularnewline
\hline 
8.8 & 8 & 0-100 & 0.0009-0.023\tabularnewline
\cline{2-4} 
 & 90 & 0-200 & 0.01-0.048\tabularnewline
\hline
\end{tabular}
\par\end{centering}

\caption{Ranges of $\tau_{eff}$ values for kinematical regions shown in the
figures.\label{tab:taueff}}
\end{table}

We will argue in Sec.~\ref{sec:NumericalPredictions} that the behavior of the nuclear
correction to the $Z/\gamma^*$ cross section 
can be understood qualitatively from the plots of the nuclear
correction to the PDFs in Figs.~\ref{pdfratio8} and \ref{pdfratio90}, taking the $x_{1c}$ 
and $x_{2c}$ as the reference momentum fractions. 
The relation is accurate within 5-10\% and works better when small $\xi$ values dominate, 
since $h(\xi_1,\xi_2)$ in Eq.~(\ref{goodlim}) is a monotonous function for small $\xi_{1,2}$. 
More detailed predictions require to know 
the behavior of $h(\xi_1,\xi_2)$ along the full integration range 
in Eq.~(\ref{goodlim}). For the bulk of $Z$ bosons, $Q_T$ is much smaller than $Q$, and
Eq.~(\ref{typx}) reduces to the familiar 
\be
\xi_{1,2} \approx x_{1,2} = (Q/\sqrt{S}) e^{\pm y}.
\ee
However, Eq.~(\ref{typx}) also clarifies the non-trivial $Q_T$
dependence of our numerical results, to which we turn next.

\section{Numerical predictions}
\label{sec:NumericalPredictions}

\subsection{Setup}

The FGS PDFs were implemented in the CSS resummation codes \textsc{Legacy} 
and \textsc{ResBos} \cite{Balazs:1997xd,Landry:2002ix} and used
to compute Drell-Yan pair production cross sections in two
characteristic mass ranges of $Q=70-110$ and $5-20$ GeV and for the boson rapidity
$|y| < 3.5$ corresponding to $x> 10^{-5}$ at all considered energies.
\textsc{ResBos} performs 
resummation of the initial-state QCD radiation and decay of massive bosons
along the lines described in Sec.~\ref{sec:QTResum}.  At
$Q\approx M_Z$, we computed  the dominant process of resonant $Z$ boson contribution, 
as well as small contributions from the continuous $\gamma^*$
background and $Z-\gamma^*$ interference.
At $Q=5-20$ GeV, we included the virtual photon contribution only, as the $Z$-mediated
cross section is negligible in this mass range.

Nuclear PDFs are generated according to Eq.~(\ref{pdfratio})
using a CT10 NLO family of PDFs of the proton \cite{Lai:2010vv}. 
To evaluate the nuclear
correction to the differential cross section, we introduce the ratio
\ba
r_{\sigma}(v)\equiv \frac{(\frac{d\sigma(A+B\rightarrow ZX)}{dv})_{with\ nucl. corr.}}
{(\frac{d\sigma(A+B\rightarrow ZX)}{dv})_{no\ nucl. corr.}},
\label{rsigma}
\ea
where $v=y$ or $Q_T$, in select ranges of $Q$ and/or $y$. The denominator 
is calculated assuming the absence of nuclear modifications of nuclear PDFs, {\it i.e.},
according to Eq.~(\ref{pdfratio}) with $R_{j}(x,Q^{2})=1$.
The cross sections are computed for the following values of the center-of-mass energy at the LHC:
$\sqrt{S_{pA}}=4.4$ TeV and 8.8 TeV in the proton-lead scattering and $\sqrt{S_{AA}}=2.76$ TeV 
and 5.52 TeV in the lead-lead scattering.

In the perturbative resummed form factor (\ref{Wpert}), the functions 
${\cal A},$ ${\cal B}$, and ${\cal C}_{j/a}$ were evaluated 
to orders $\alpha_{s}^{3},$ $\alpha_{s}^{2}$,
and $\alpha_{s}$, respectively \cite{Collins:1984kg,Davies:1984hs,Davies:1984sp,Moch:2004pa}.
The scale parameters in $\widetilde{W}^{pert}$
were chosen as $\{C_1=C_3=2b_0,C_2=2\}$. The nonperturbative
contributions were introduced according to Ref.~\cite{GNW} 
using the $b_*$ convention with $b_{max} =1.5 \mbox{ GeV}^{-1}$ and 
$\widetilde W^{NP}(b)=\exp\left[-a\, b^{2}\right]$. We take $a=1.1\mbox{
  GeV}^2$ by default in $Z$ production and $a=0.3$ or $1.1\mbox{
  GeV}^2$ in low-$Q$ Drell-Yan production. 
We assume the same $a$ value in the numerator and denominator 
of the ratio $r_\sigma$, so that the dependence on $a$ 
cancels well inside $r_\sigma$. The predictions for $r_\sigma$ that
will be shown are hardly sensitive to the value of $a$ that is assumed.

The $Y$ term was estimated to NLO in photon-mediated
subprocesses, and to NNLO in the pure $Z$ cross section by using the 
two-loop correction from Ref.~\cite{Arnold:1988dp}. The
renormalization and factorization scales were set to $2Q$ in the $Y$
piece. All these settings are in good agreement with 
ATLAS $Q_T$ distributions for $pp\rightarrow Z/\gamma^* X$
\cite{Aad:2011gj}, as has been found in Ref.~\cite{GNW}.

\subsection{Proton-lead collisions}

The $r_\sigma$ ratios for proton-lead collisions are shown in
Figs.~\ref{fig:pAZy}-\ref{fig:pAZQT2} for $Z$ boson production in a
representative interval $70 < Q < 110$ GeV 
and in Figs.~\ref{fig:pAgy}-\ref{fig:pAgQT2} for low-$Q$ Drell-Yan
process at $5<Q<20$ GeV. 
The dashed purple line and short-dashed black line were computed using the 
FGS10\_H and FGS10\_L nuclear PDFs, respectively.

\subsubsection{$Z$ pole region \label{sec:pAZ}}

\begin{figure}[ht]
\begin{center}
\includegraphics[width=1.0\textwidth]{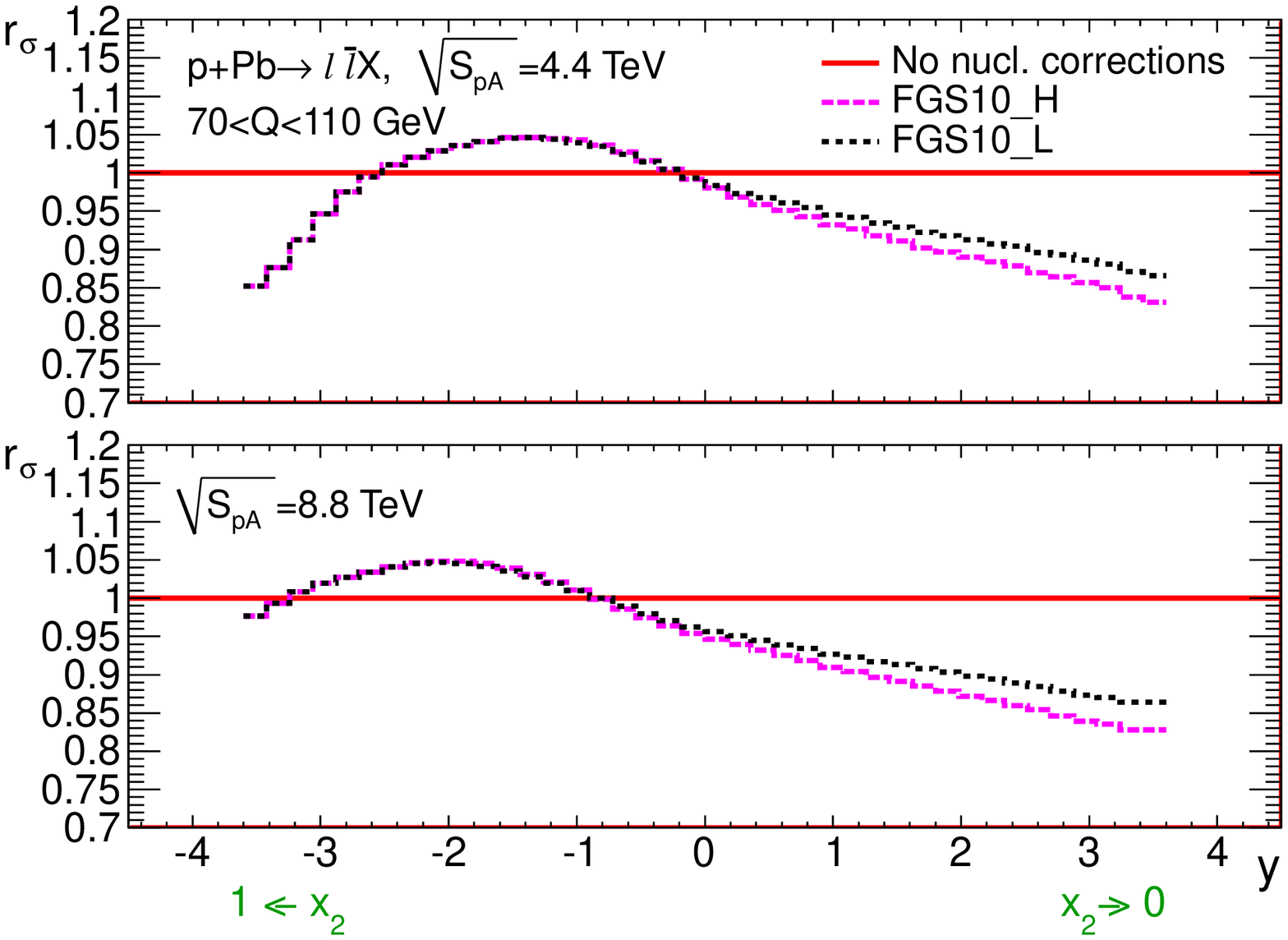}
\caption{The nuclear correction ratio $r_{\sigma}(y)$, as defined in Eq.~(\ref{rsigma}), plotted 
for $70<Q<110$ GeV vs. the lepton pair rapidity $y$ in proton-lead collisions.}

\label{fig:pAZy}
\end{center}
\end{figure}

We start with the plots of the $r_\sigma$ ratio for the $Z$ rapidity distribution 
in Fig.~\ref{fig:pAZy}, as they are the simplest. In
$pA$ collisions, the nuclear  
correction acts on the PDF of the lead nucleus but not on the proton PDF.
The nuclear correction depends on $\xi_2$, which is of order
$x_{2c} = \tau_{eff}\, \exp(-y)$ for the most part. We can therefore understand 
the overall magnitude of the nuclear correction within 5-10\% 
by estimating $x_{2c}$ for the given $y$ and $Q_T$, as has been argued 
in Sec.~\ref{sec:exptests}.
 
Since $Q_T$ is much smaller than $Q$ for the majority of $Z$ events,
a scan of $r_\sigma(y)$ over $y$  in Fig.~\ref{fig:pAZy}  essentially 
translates into a scan over the momentum fraction $x$ in the PDF 
nuclear correction in Fig.~\ref{pdfratio90}. Negative (positive) values of $y$ 
correspond to large (small) values of
$x=(Q/\sqrt{S_{pA}})\exp(-y)$. The typical $x$ value (equated to
$x_{2c}$) can be found for each $y$ from Fig.~\ref{kine}
and Table~\ref{tab:taueff}. 

In Fig.~\ref{pdfratio90} we see the small-$x$ shadowing 
at $x\lesssim 2\cdot 10^{-2}$, antishadowing at
$2\cdot 10^{-2} \lesssim x \lesssim 0.3$, and large-$x$ suppression at
$x \gtrsim 0.3$. The exact boundaries between the three regions depend on the parton flavor.
In the upper inset of Fig.~\ref{fig:pAZy}, these regions translate
into $r_\sigma < 1$ at $y >0$, $r_\sigma > 1$ at $-2.6 < y < 0$, and
$r_\sigma < 1$ for $y < -2.6$. The suppression is the most
pronounced (of order 15-20\%) in the forward rapidity regions.  
The FGS10\_L and FGS10\_H predictions differ in the
small-$x$/large-$y$ shadowing region, but are very close otherwise. 

In the lower inset (obtained for $\sqrt{S_{pA}}=8.8$ TeV), 
we observe a qualitatively similar behavior of $r_\sigma$ as in the
upper inset (for $\sqrt{S_{pA}}=4.4$ TeV), but  
the antishadowing ``hill'' is shifted toward lower $y$ values, $-3.2
\lesssim y < -0.8$,  
as a result of reduced $\tau_{eff}$ for the same $Q$. 

The measurement of $Q_T$ of the lepton pair provides an additional
handle for probing the $x$ dependence. For $y$ fixed,  
a larger $Q_T$ value leads to a larger $\tau_{eff}$ and $x_{2c}$ 
as compared to $Q_T = 0$. Ranges  
of $\tau_{eff}$ in the $Q_T$ intervals shown in the figures are listed in 
Table~\ref{tab:taueff}.

We also observe that the flavor dependence of scattering contributions is
distinct at small and large $Q_T$.  At $Q_T \ll Q$, 
the scattering proceeds largely through $q\bar q$ annihilation, 
while at $Q_T \approx Q$ the 
$qg$ Compton scattering becomes competitive. The analysis of 
$Q_T$ dependence can thus identify nuclear corrections attributed 
to the gluon PDF.

\begin{figure}[ht]
\begin{center}
\includegraphics[width=1.0\textwidth]{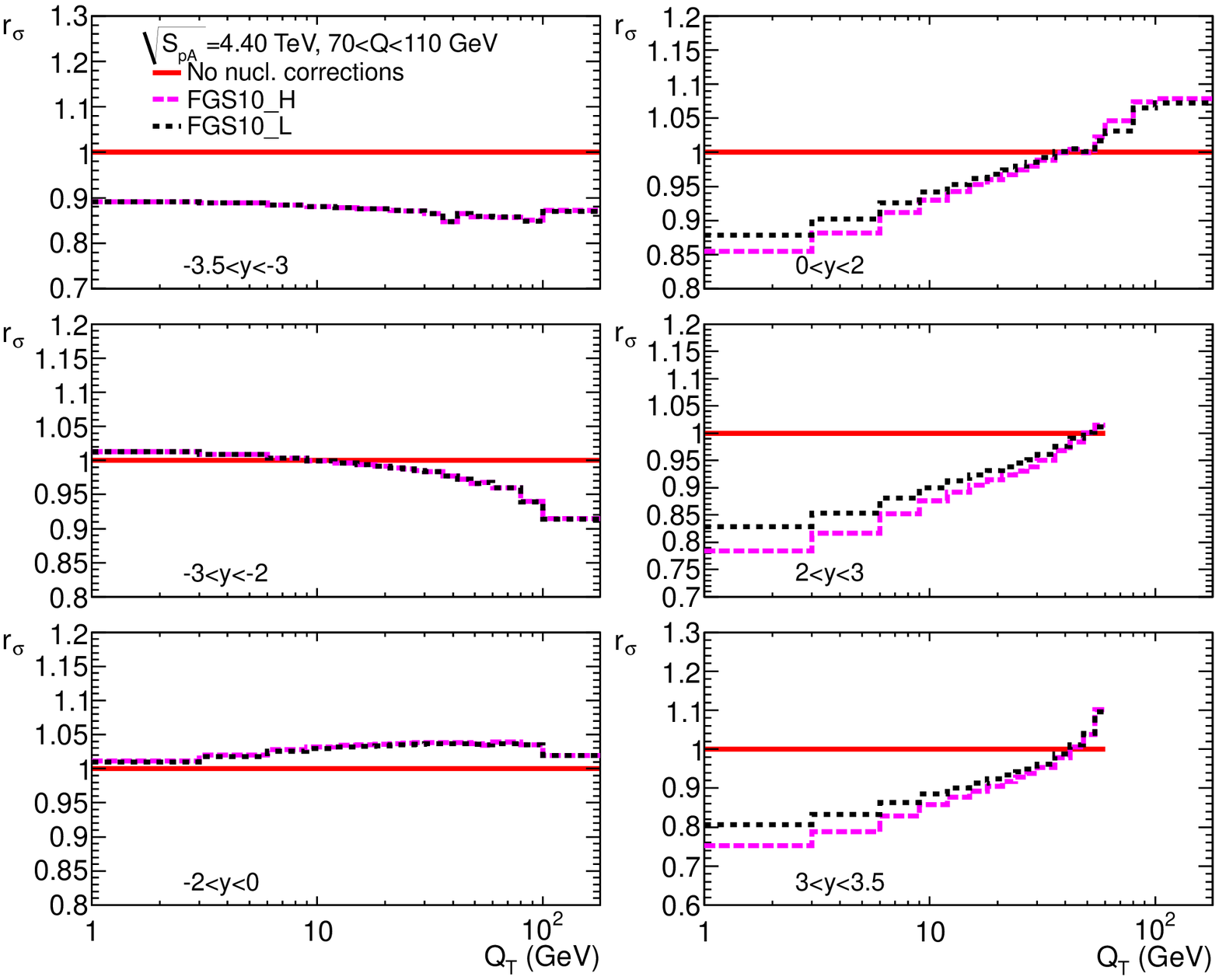}
\caption{The nuclear correction ratio $r_{\sigma}(Q_T)$ for the $Q_{T}$ distribution and $70<Q<110$ GeV
in proton-lead collisions at $\sqrt{S_{pA}}=4.4$ TeV, in six bins of $y$.}
\label{fig:pAZQT1}
\end{center}
\end{figure}

\begin{figure}[ht]
\begin{center}
\includegraphics[width=1.0\textwidth]{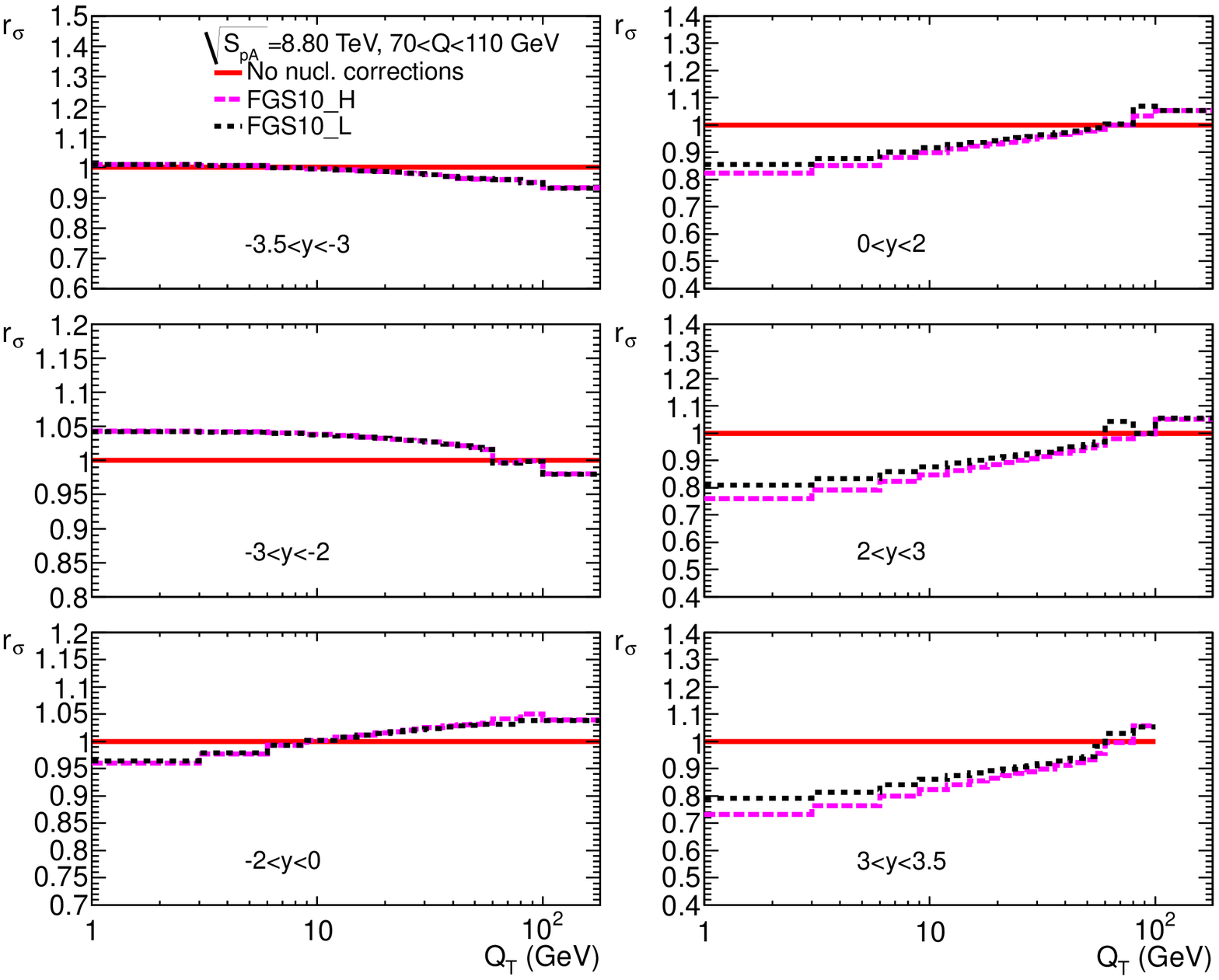}
\caption{Same as Fig.~\ref{fig:pAZQT1}, for $\sqrt{S_{pA}}=8.8$ TeV.}
\label{fig:pAZQT2}
\end{center}
\end{figure}

The $Q_T$ dependent ratios $r_\sigma(Q_T)$ for $pA\rightarrow Z$
production are plotted at 4.4 and 8.8 TeV
in Figs.~\ref{fig:pAZQT1} and \ref{fig:pAZQT2} in the specified bins of $y$. 
The $Q_T$ bins were chosen as in the $pp\rightarrow ZX$ measurements
by ATLAS \cite{Aad:2011gj}.  
The rapidity dependence is not symmetric with respect to $y=0$
in the $pA$ case. Hence 
the $Q_T$ distributions are plotted separately for positive and negative $y$. 

In both figures, the value of $r_\sigma$ at low $Q_T$ in each bin is
in rough correspondence with the $r_\sigma(y)$ value for  
the corresponding $y$ in the rapidity distribution. This is expected,
since most $Z$ events that contribute to the rapidity distribution  
have small $Q_T$. As $Q_T$ increases, the $r_\sigma(Q_T)$ changes
non-trivially, especiall in the positive $y$ bins  
(right columns in the figures), where significant small-$x$ shadowing
(of order 15-20\%) at $Q_T\rightarrow 0$ is superceded  
by antishadowing of up to 10\% at $Q_T > 50-100$ GeV.

\subsubsection{Low-$Q$ region \label{sec:pAg}}
\begin{figure}[ht]
\begin{center}
\includegraphics[width=1.0\textwidth]{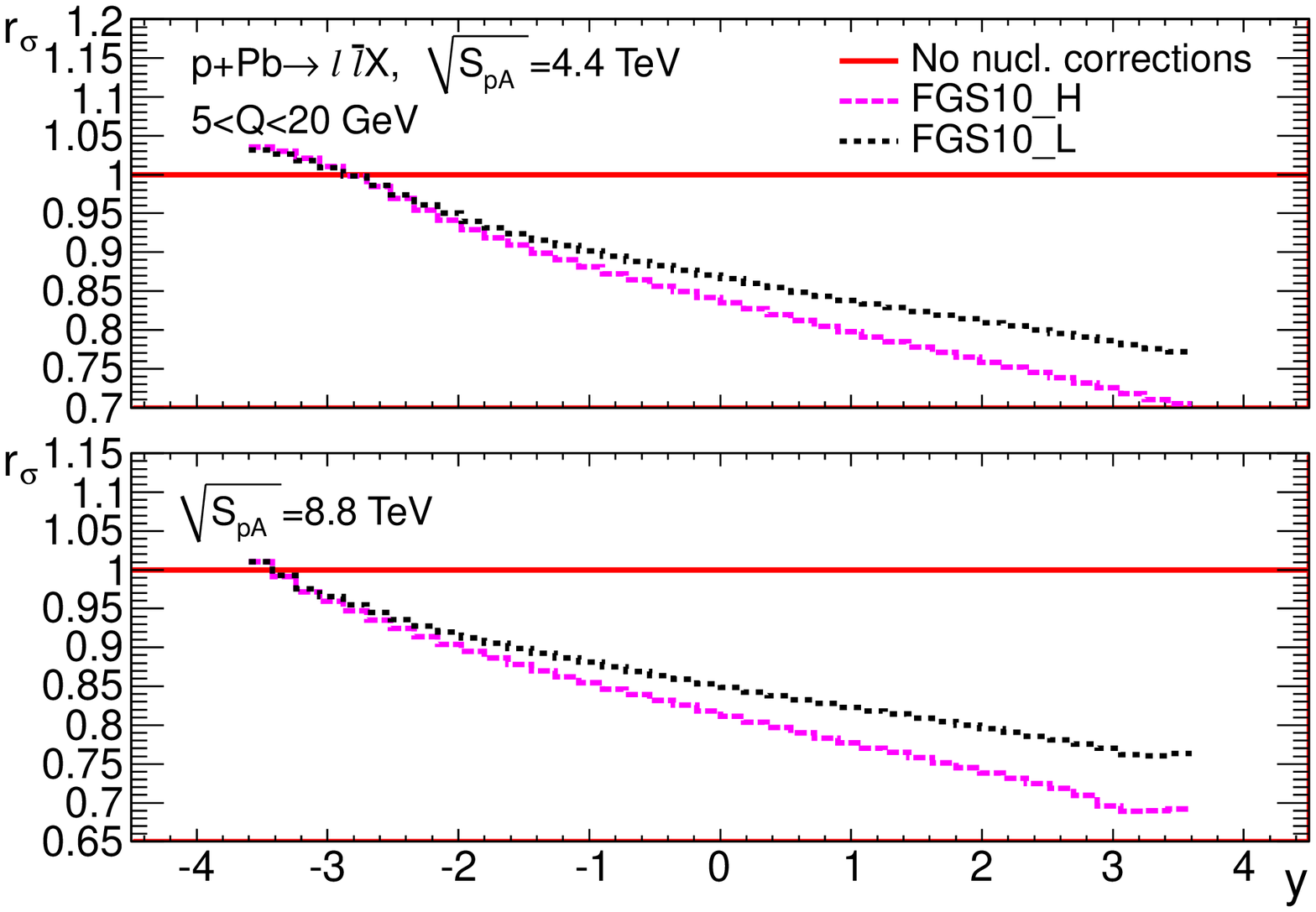}
\caption{The nuclear correction ratio $r_{\sigma}(y)$, as defined in Eq.~(\ref{rsigma}), plotted 
for $5<Q<20$ GeV vs. the lepton pair rapidity $y$ in proton-lead collisions.}
\label{fig:pAgy}
\end{center}
\end{figure}

In the low invariant mass range, $5 < Q < 20$ GeV,
$Z$-mediated $e^+e^-$ production is negligible compared to
virtual photon contributions.  In this case Table~\ref{tab:taueff} and Fig.~\ref{kine}
tell us that the Drell-Yan process probes 
much smaller values of $x_{2,c}$, of order a few $10^{-3}$.
The $r_\sigma$ rapidity distribution in Fig.~\ref{fig:pAgy} shows
that small-$x$ nuclear shadowing dominates 
across the full rapidity range, with the exception of 
a small band of large negative rapidities below $-2.5$. The magnitude
of shadowing exhibits a monotonous increase from zero at $y\approx -2.5$ to 
25-30\% at $y \approx +3$. 

\begin{figure}[ht]
\begin{center}
\includegraphics[width=1.0\textwidth]{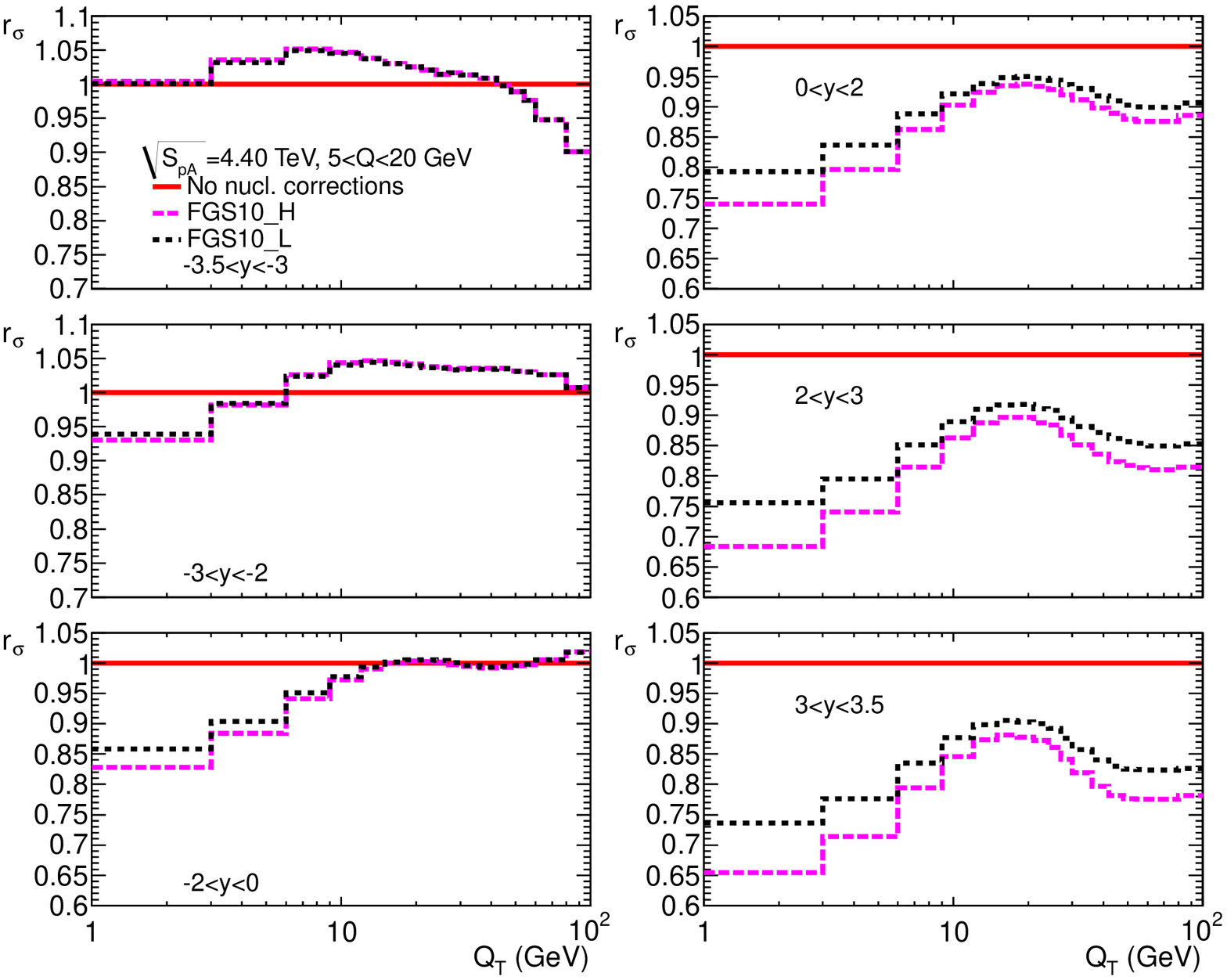}
\caption{The nuclear correction ratio $r_{\sigma}(Q_T)$ for the $Q_{T}$ distribution and $70<Q<110$ GeV
in proton-lead collisions at $\sqrt{S_{pA}}=4.4$ TeV, in six bins of $y$.}
\label{fig:pAgQT1}
\end{center}
\end{figure}

\begin{figure}[ht]
\begin{center}
\includegraphics[width=1.0\textwidth]{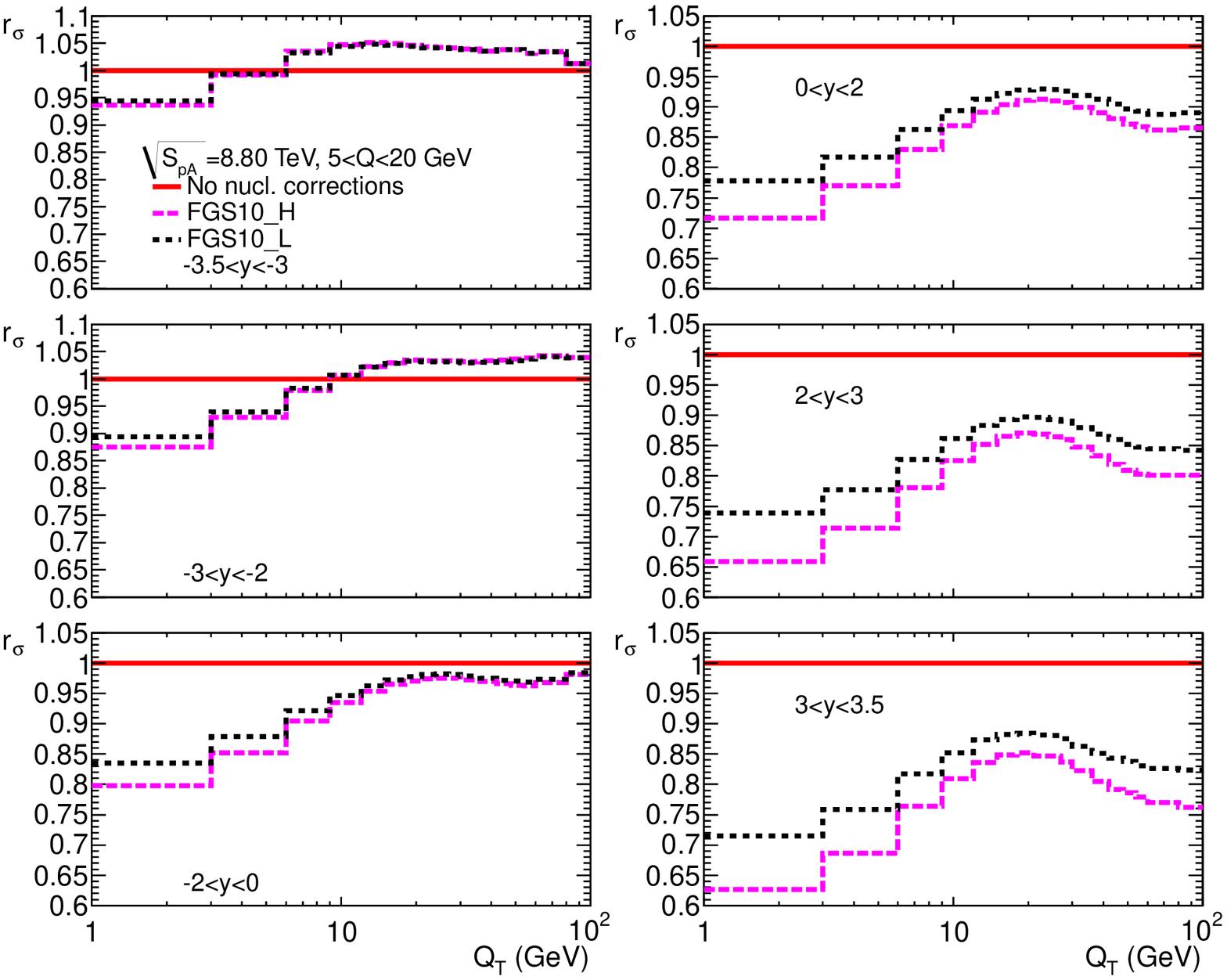}
\caption{Same as Fig.~\ref{fig:pAgQT1}, for $\sqrt{S_{pA}}=8.8$ TeV.}
\label{fig:pAgQT2}
\end{center}
\end{figure}

Even when shadowing dominates across in some $y$ interval,
antishadowing may still occur at this $y$ if $Q_T$ is high enough. This is 
illustrated by Figs.~\ref{fig:pAgQT1} and \ref{fig:pAgQT2}, 
showing $Q_T$ dependence of the $r_\sigma$ ratio in the low-mass
range. In the bins with $y < 0$ in the left columns of
Figs.~\ref{fig:pAgQT1} and \ref{fig:pAgQT2}, $r_\sigma$ is smaller
than 1 (larger than 1) at $Q_T=0$  ($Q_T=20-30$ GeV). For $y < -3$,
even large-$x$ suppression of up to 10\% 
occurs at the largest $Q_T$ values close to 100 GeV. 

At positive $y$ in Figs.~\ref{fig:pAgQT1} and \ref{fig:pAgQT2}
(right columns), small $Q_T$ shadowing is more prominent. %% another interesting feature is noticeable. 
 In this $y$ region, corresponding to $x < 2\cdot 10^{-3}$ for $Q_T=0$,
shadowing can reach 40\% in the lowest $Q_T$ bins. As the transverse
momentum increases above $Q_T = 10\mbox{ GeV}$, $r_\sigma$ remains below unity, in consistency with 
the $x_{2c}$ value that is still quite small. Detailed variations of $r_\sigma$ are now 
more intricate, as a wide range of the momentum fractions $\xi_2$ contributes at these $y$: 
$\xi_{2,min}\leq \xi_2 \leq 1$, where $\xi_{2,min}\approx x_2$ is much smaller 
than 0.01. As a result, $r_\sigma$ at $Q_T>10 $ GeV 
shows both the dependence on the small-$x$ shadowing model (from $\xi_2 < 0.01$)  
and a ``hill'' from the antishadowing and enhancements in the valence PDFs at $\xi_2>0.01$.

It can be further shown that the $qg$ Compton scattering contributes of order 80\% 
at the largest $Q_T$ shown in the figures. The Drell-Yan process at low $Q$ and large $Q_T$ is 
therefore an excellent probe of nuclear corrections to the nuclear PDFs, in full analogy 
with $pp$ case \cite{Berger:2001wr,Kang:2008wv}.

\subsection{Numerical results for lead-lead collisions}
\subsubsection{$Z$ pole region \label{sec:AAZ}}
\begin{figure}[ht]
\begin{center}
\includegraphics[width=1.0\textwidth]{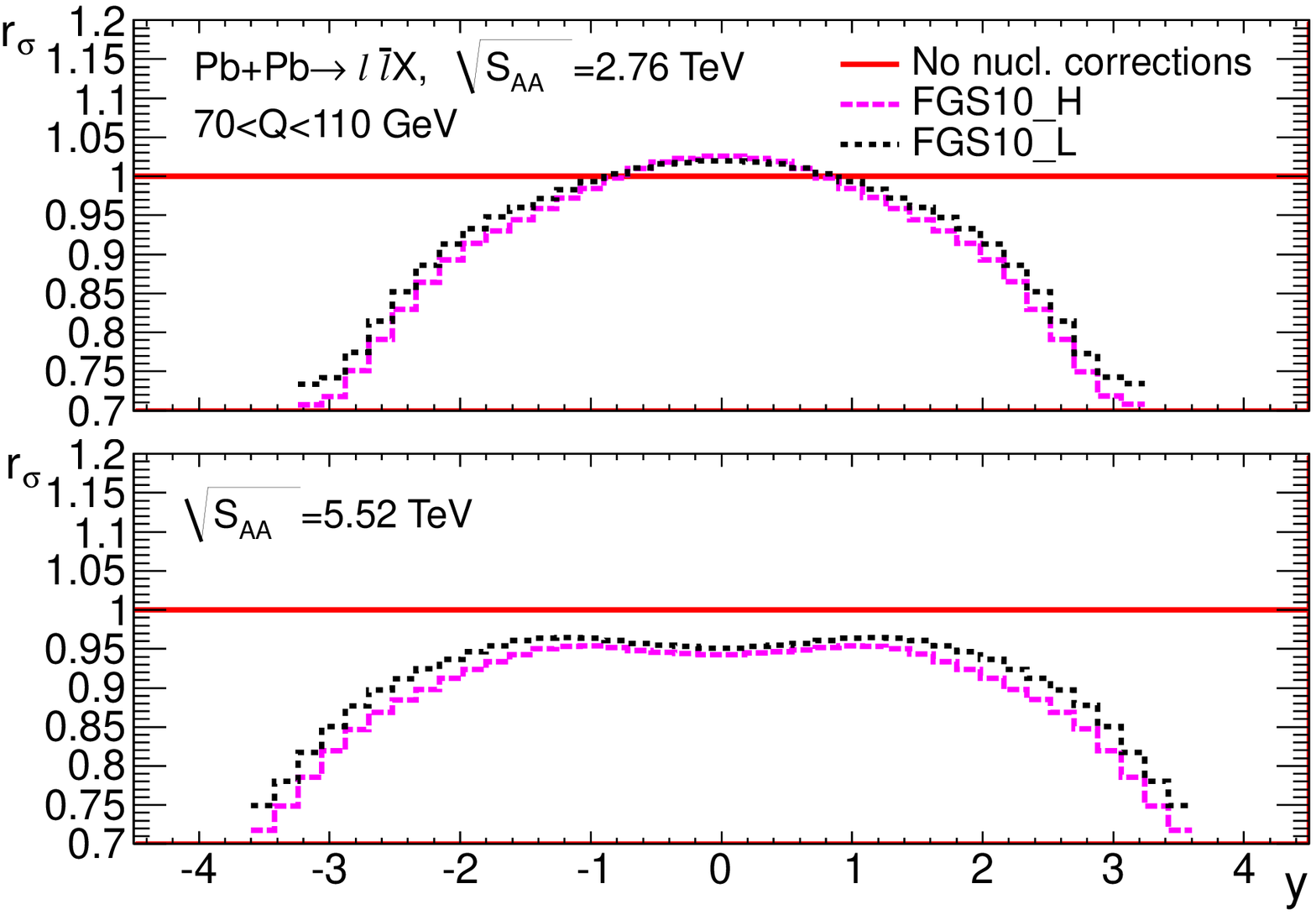}
\caption{The nuclear correction ratio $r_{\sigma}(y)$, as defined in Eq.~(\ref{rsigma}), plotted 
for $70<Q<110$ GeV vs. the lepton pair rapidity $y$ in lead-lead collisions}
\label{fig:AAZy}
\end{center}
\end{figure}

The nuclear correction is more complex
in lead-lead collisions, since both incoming beams contribute 
to the nuclear effects. In this case the rapidity distribution is
symmetric, as can be seen in Fig.~\ref{fig:AAZy}. The 
$\tau_{eff}$ values with $Q_T=0$ are
equal approximately to 0.033 at $\sqrt{S_{AA}}=2.76$ TeV
and 0.016 at $\sqrt{S_{AA}}=5.52$ TeV and correspond to mild antishadowing
and small-$x$ shadowing, respectively. Consequently the $y=0$ region
in Fig.~\ref{fig:AAZy} exhibits antishadowing of 2-3\% for
$\sqrt{S_{AA}}=2.76$ TeV (upper inset) and 5\% shadowing for
$\sqrt{S_{AA}}=5.52$ TeV (lower inset). In the central region,
predictions based on two nuclear models are almost indistinguishable.

As $|y|$ increases, the strong small-$x$ shadowing in one of the
initial-state nuclei overtakes the mild increase in $r_\sigma$ due to
antishadowing associated with the second nuclei. At the largest
rapidity shown ($|y| \rightarrow 3.5$) the shadowing reaches up to 30\% and
has different magnitude in the two nuclear models.

\begin{figure}[ht]
\begin{center}
%\hspace{-1cm}
\includegraphics[width=0.8\textwidth]{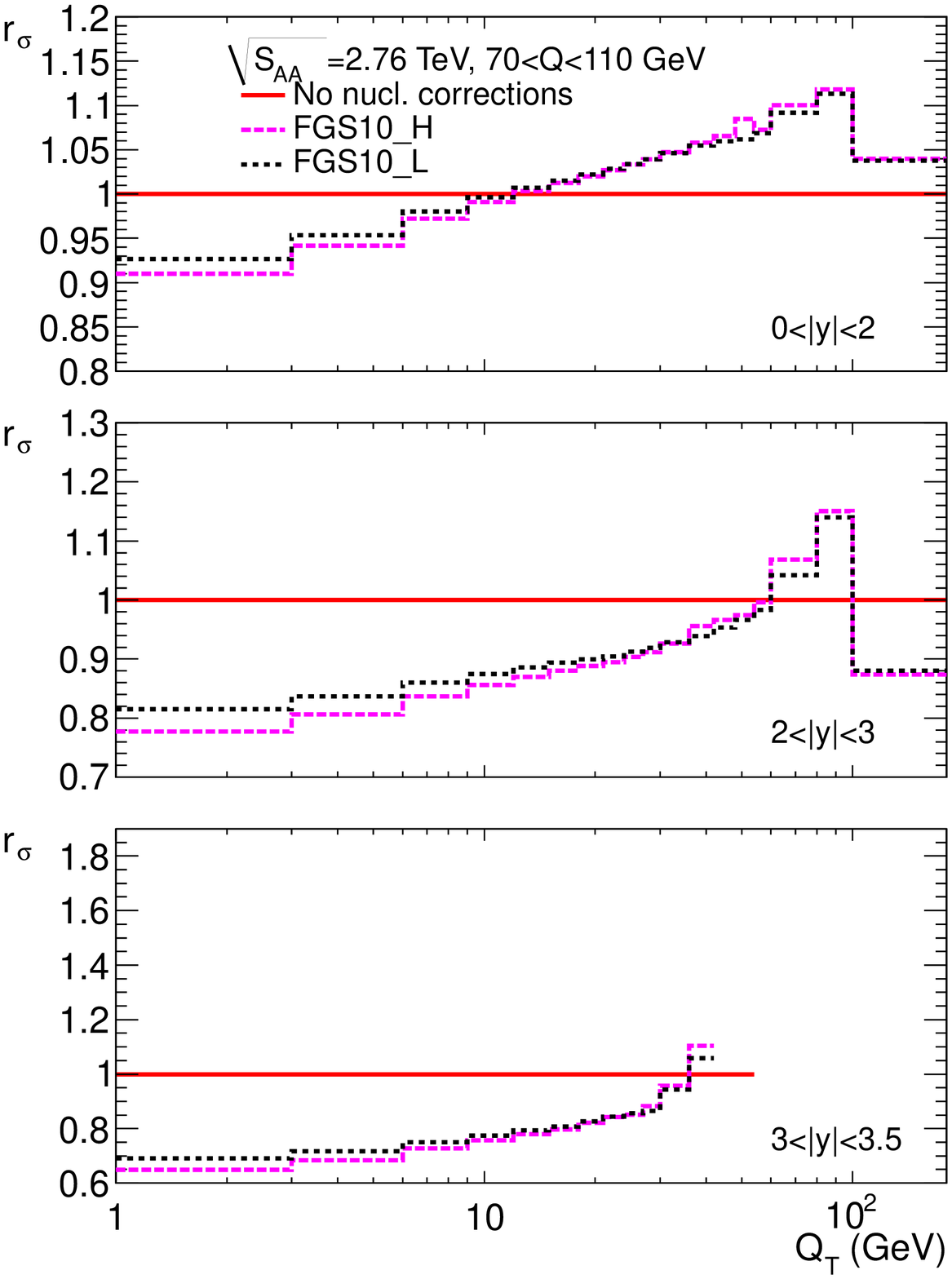}
\caption{
The nuclear correction ratio $r_{\sigma}(Q_T)$ for the $Q_{T}$ distribution and $70<Q<110$ GeV
in proton-lead collisions at $\sqrt{S_{AA}}=2.76$ TeV, in three bins of $|y|$.}
\label{fig:AAZQT1}
\end{center}
\end{figure}

\begin{figure}[ht]
\begin{center}
%\hspace{-1cm}
\includegraphics[width=0.8\textwidth]{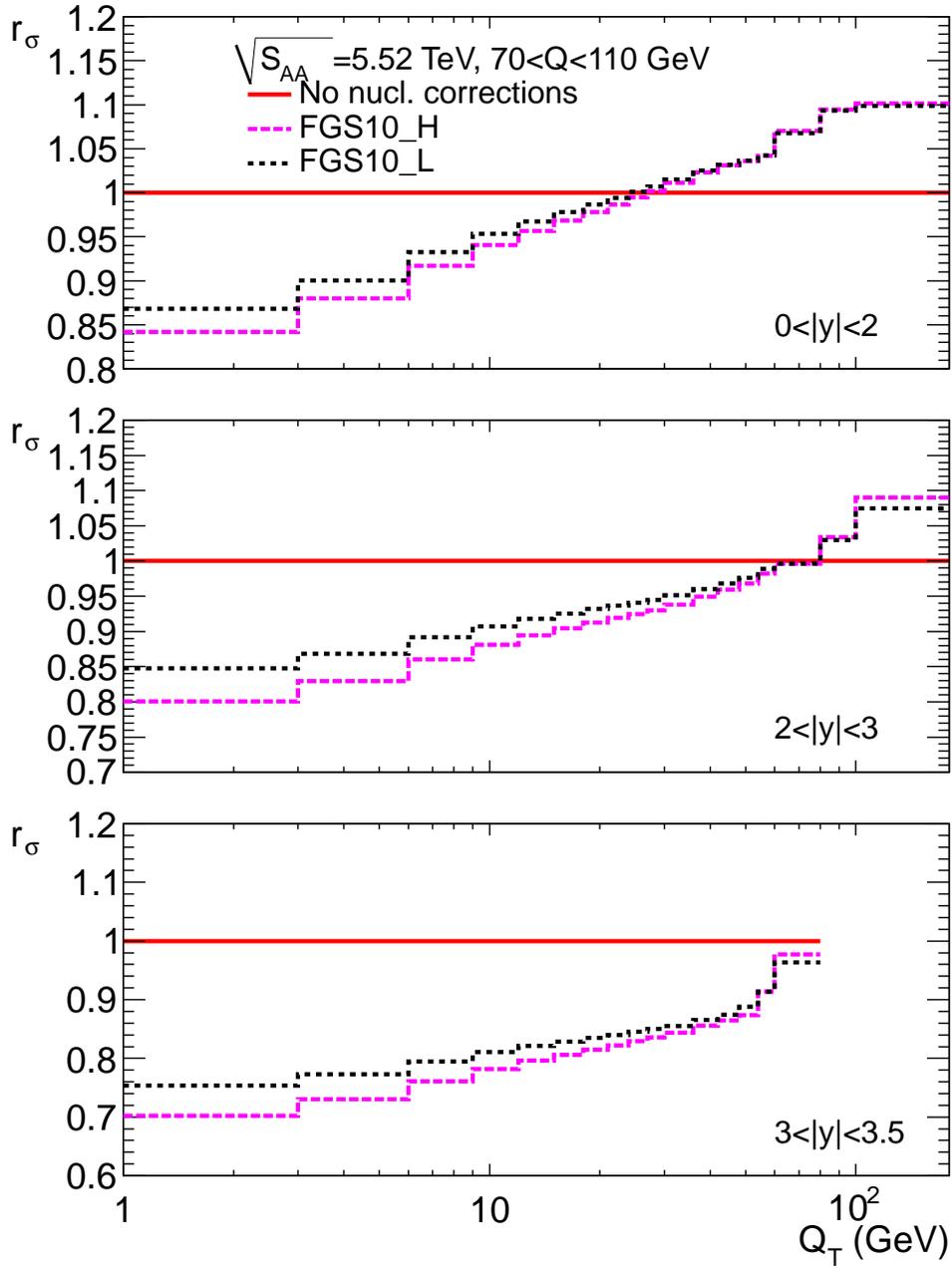}
\caption{Same as Fig.~\ref{fig:AAZQT1}, for $\sqrt{S_{PbPb}}=5.52$ TeV.}
\label{fig:AAZQT2}
\end{center}
\end{figure}

$Q_T$ dependence of $r_\sigma$ is illustrated in
Figs.~\ref{fig:AAZQT1} and \ref{fig:AAZQT2}, in three bins of $|y|$. 
We observe the strong small-$x$
shadowing (15-20\%) at the smallest $Q_T$ values, which is replaced by
antishadowing (up to 10\%) at intermidiate $Q_T$, and eventually by
large-$x$ suppression at the highest $Q_T$ attainable. The shadowing is
generally stronger for $\sqrt{S_{AA}}=5.52$ TeV (Fig.~\ref{fig:AAZQT2}) as a
result of smaller typical $\tau_{eff}$, of about 0.001 for $Q_T=0$ and
0.036 for $Q_T=100 $ GeV. 

\subsubsection{Low-$Q$ region \label{sec:AAg}}
\begin{figure}[ht]
\begin{center}
\includegraphics[width=1.0\textwidth]{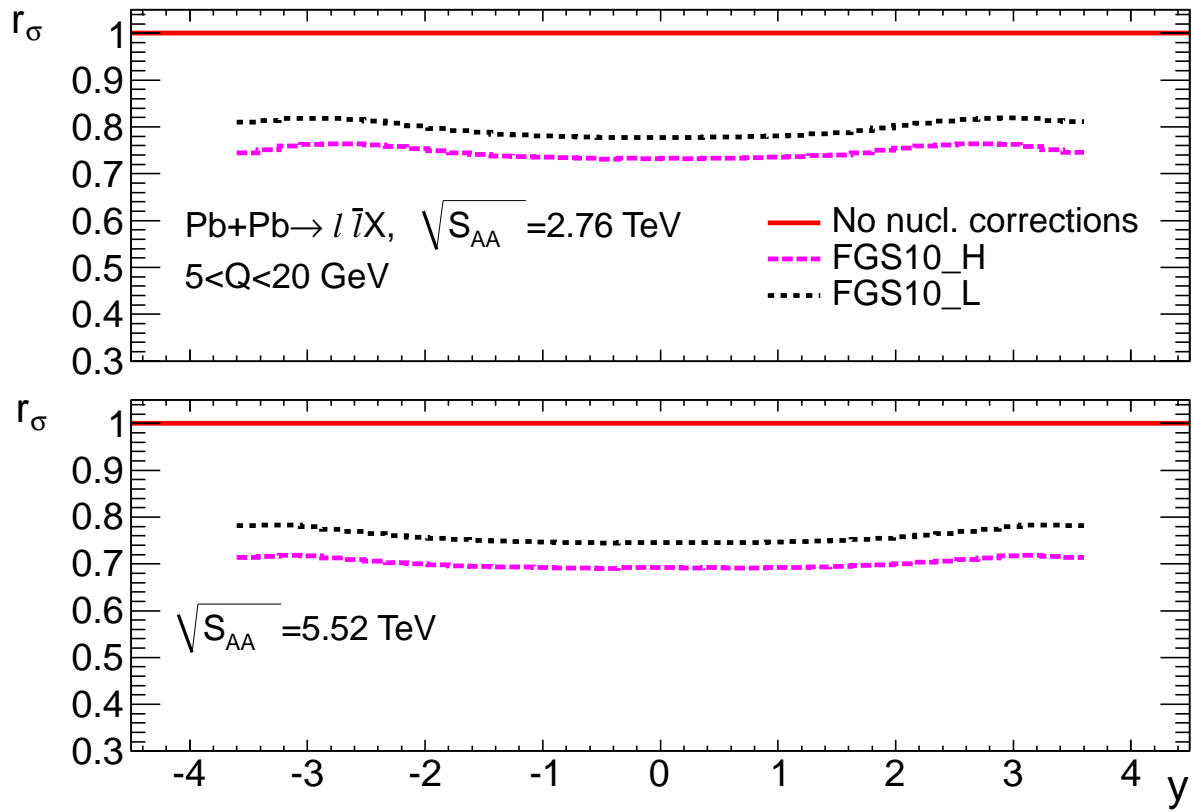}
\caption{The nuclear correction ratio $r_{\sigma}(y)$, as defined in Eq.~(\ref{rsigma}), plotted 
for $5<Q<20$ GeV vs. $y$ in lead-lead collisions}

\label{fig:AAgy}
\end{center}
\end{figure}

\begin{figure}[ht]
\begin{center}
%\hspace{-1cm}
\includegraphics[width=0.8\textwidth]{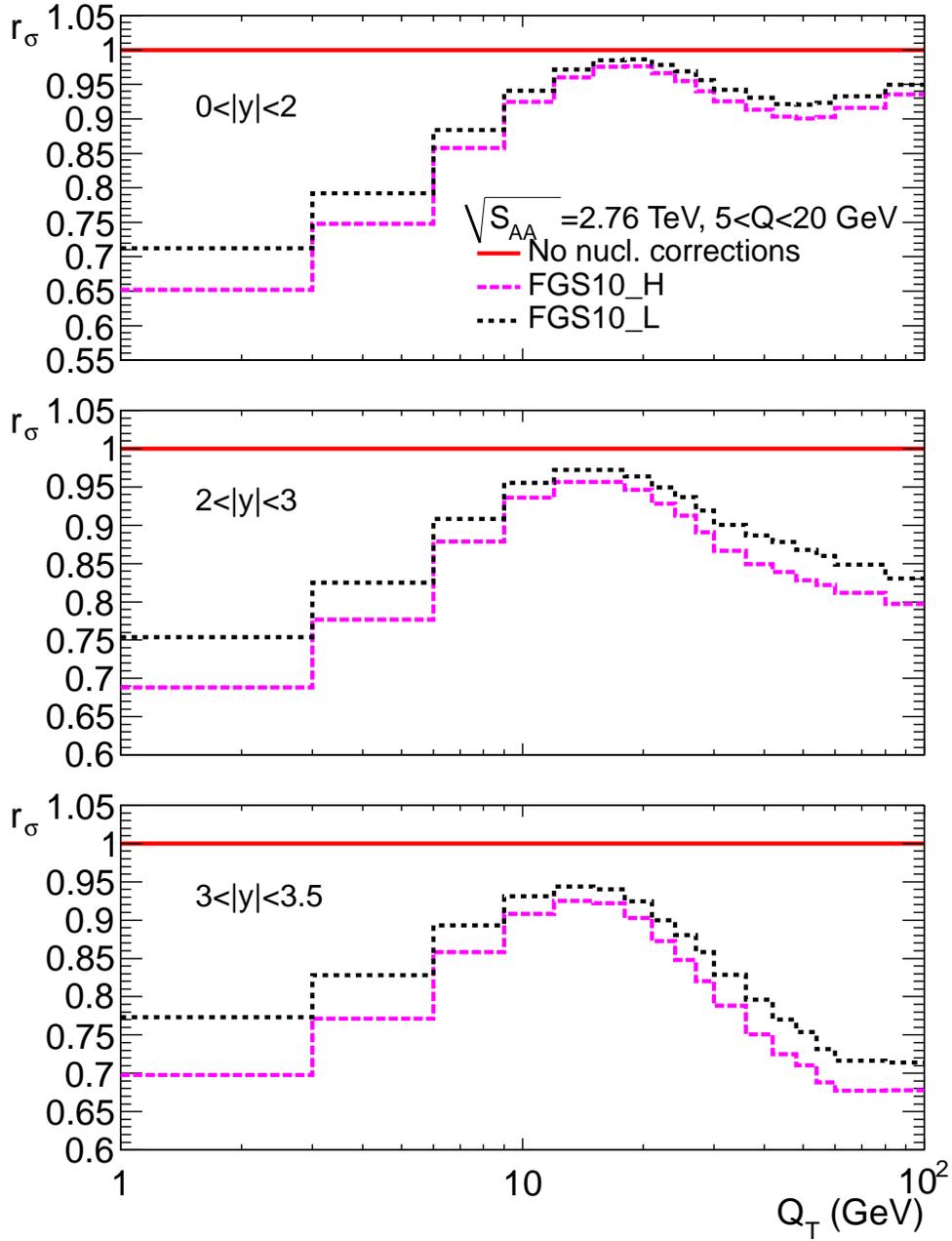}
\caption{The nuclear correction ratio $r_{\sigma}(Q_T)$ for the $Q_{T}$ distribution and $5<Q<20$ GeV
in lead-lead collisions at $\sqrt{S_{AA}}=2.76$ TeV, in three bins of $|y|$.}
\label{fig:AAgQT1}
\end{center}
\end{figure}

\begin{figure}[ht]
\begin{center}
%\hspace{-1cm}
\includegraphics[width=0.8\textwidth]{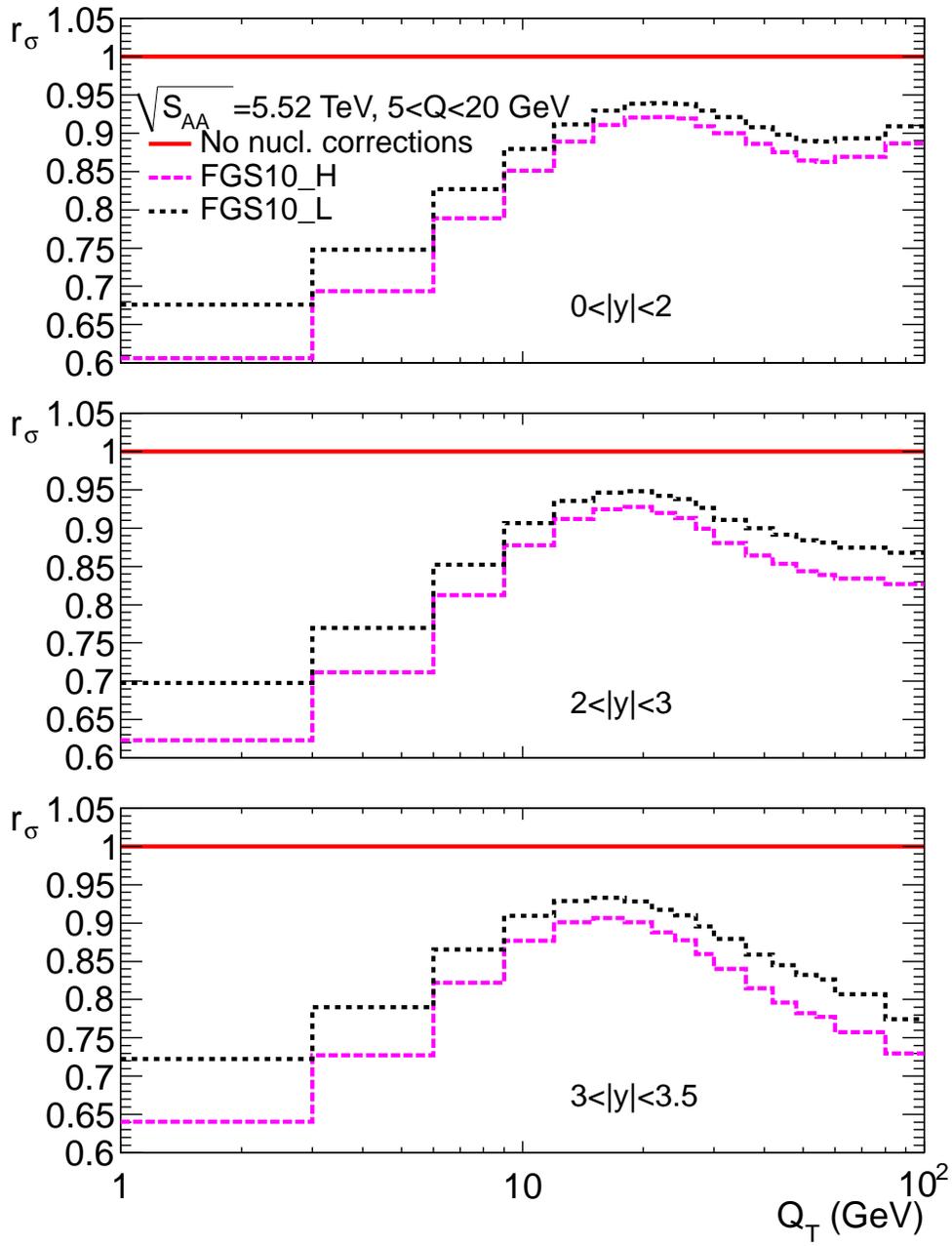}
\caption{Same as Fig.~\ref{fig:AAgQT1}, for $\sqrt{S_{AA}}=5.52$ TeV.}
\label{fig:AAgQT2}
\end{center}
\end{figure}

In the low invariant mass regime illustrated by 
Figs.~\ref{fig:AAgy}-\ref{fig:AAgQT2}, the shadowing is most
pronounced; antishadowing in one nuclei never overcomes shadowing 
in the counterpart nuclei in either $y$ distribution or $Q_T$ distribution.
Antishadowing produces characteristic bumplike features
in the lower inset of the $y$ distribution in Fig.~\ref{fig:AAgy} and
$Q_T$ distributions in Figs.~\ref{fig:AAgQT1} and \ref{fig:AAgQT2},
but never overcomes the overall suppression that results in $r_\sigma
< 1$. In the $Q_T$ distribution, the shadowing is generally stronger
at the smallest $Q_T$ (up to 40\%) than in the $y$ distribution
(25-30\%). The profile of $r_\sigma$ at the intermediate and large $Q_T$ value again
displays a ``hill'', similarly to the $pA$ case.

\subsection{Nonperturbative smearing of the transverse momentum distribution}
One of the key uncertainties in determination of the nuclear PDF
corrections from Drell-Yan $Q_T$ distributions is associated with
the small-$Q_T$ nonperturbative function $\widetilde{W}^{NP}(b,Q)$ that is 
not completely known in nuclear scattering. In our analysis, $\widetilde{W}^{NP}(b,Q)$ 
has been determined from $pp$ collisions (cf. Sec.~\ref{sec:QTResum}) and parametrized 
as 
\be
\widetilde{W}^{NP}(b,Q) = \exp\left(-a(Q) b^2 \right),
\ee
with $a(Q)=1.1$ GeV$^2$ at $Q\approx M_Z$ and $a=0.3-1.1\mbox{ GeV}^2$
at low $Q$. In the previous subsections, the ratio $r_\sigma$ has quantified
the effect of the nuclear correction to the nuclear PDF and 
was constructed to be almost independent of $a$. But for the actual $Q_T$ distributions,
the nuclear modifications may also arise from the nonperturbative function 
$\widetilde W_{NP}(b,Q)$, which has impact on production of Drell-Yan pairs at $Q_T < 20$ GeV.

Analysis of the experimental data on Drell-Yan dilepton production in $pA$ 
collisions at $\sqrt{S}=40\mbox{ GeV}$  finds  \cite{Johnson:2006wi} 
a very small $A$-dependent broadening of the $Q_T$ distribution. 
At this energy (corresponding to $x\gtrsim 0.1$) and $Q_T < 3\mbox{ GeV}$, 
$\Delta \langle Q_T^2\rangle$ grows approximately linearly with
$A^{1/3}$ and reaches 0.1 GeV$^2$ for the heaviest nuclei. 
Moreover the analysis indicates that the data at lower energies are consistent 
with the energy-independent $\Delta\langle Q_T^2 \rangle$.
Physically, the broadening of the $Q_T$ 
distribution can be interpreted as multiple rescattering of the
incoming quark off $A^{1/3}$ nucleons. At smaller momentum fractions
$\xi$, the broadening from rescattering may be larger due to
the increase of the gluon density at small $x$. At the same time the $A$
dependence of the broadening effect may become weaker due to the leading-twist
nuclear shadowing.    

In the CSS resummation formalism, broadening is accommodated by
modifying the nonperturbative $k_T$ distribution in the
nucleus at small $x$ in the region of nuclear shadowing. 
The nuclear shadowing is stronger for small virtualities, 
so that effective $k_T^2$ at scale 1 GeV may be somewhat  larger than 
in the nucleon case. 

To account for this possibility, and the fact that $a$ has uncertainties even 
in $pp$ collisions, we give the plots of $d\sigma/dQ_T$ 
for a rather wide range  of $a$  in a representative case of 
lead-lead collisions at $\sqrt{S_{AA}}=2.76$ TeV in the
shadowing model FGS10\_H. 

\begin{figure}[ht]
\begin{center}
\includegraphics[width=0.7\textwidth]{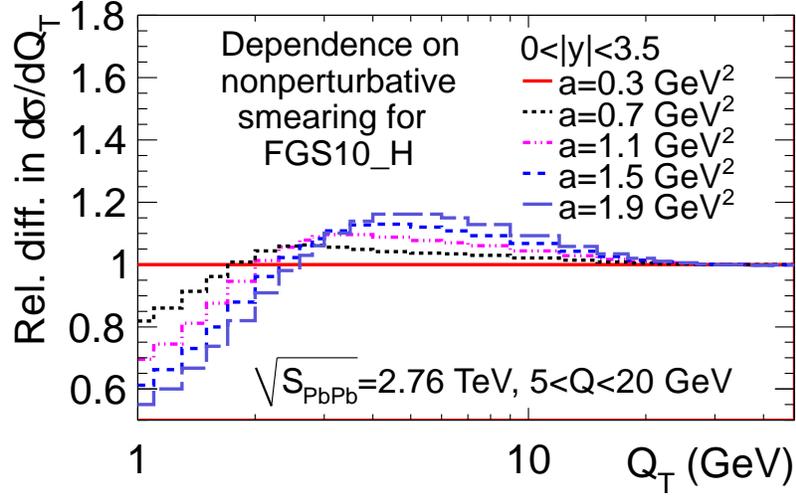}
\caption{Dependence on nonperturbative smearing of the $Q_{T}$ distribution in lead-lead collisions 
for $5 < Q < 20$ GeV at $\sqrt{S_{AA}}=2.76$ TeV. }
\label{fig:adepQ}
\end{center}
\end{figure}

In Fig.~\ref{fig:adepQ} we compare predictions 
for $d\sigma/dQ_T$ at $5 < Q < 20\mbox{ GeV}$ 
for the range $a=0.3, 0.7, 1.1., 1.5,$ and $1.9\mbox{ GeV}^2$, divided by 
$d\sigma/dQ_T$ for $a=0.3\mbox{ GeV}^2$. The value of $a=0.3\mbox{ GeV}^2$ is 
of order of the average $a$ value in $pp$ collisions 
in this $Q$ interval in fixed-target (large-$x$) Drell-Yan production
\cite{Konychev:2005iy}. Larger $a$ values would reflect
the increased nonperturbative $k_T$ in the nuclei
and result in broader $Q_T$ distributions that may be more typical 
for small $x$ values \cite{Berge:2004nt} and/or nuclear targets. 
In this specific example, increasing $a$ from $0.3$ 
to $1.9\mbox{ GeV}^2$ increases the average $\langle Q_T^2\rangle$ in the affected 
interval $0 < Q_T < 20$ GeV by $2.5\mbox{ GeV}^2$ 
(from $28.6$ to $31.1\mbox{ GeV}^2$, which is much larger than $\Delta\langle Q_T^2 \rangle$ in a 
fixed-target experiment discussed above). The dependence of $\langle Q_T^2 \rangle$ 
on $a$ reflects the interplay of perturbative and nonperturbative QCD contributions
and varies with $\sqrt{S}$ and $Q$.

\begin{figure}[ht]
\begin{center}
\includegraphics[width=0.7\textwidth]{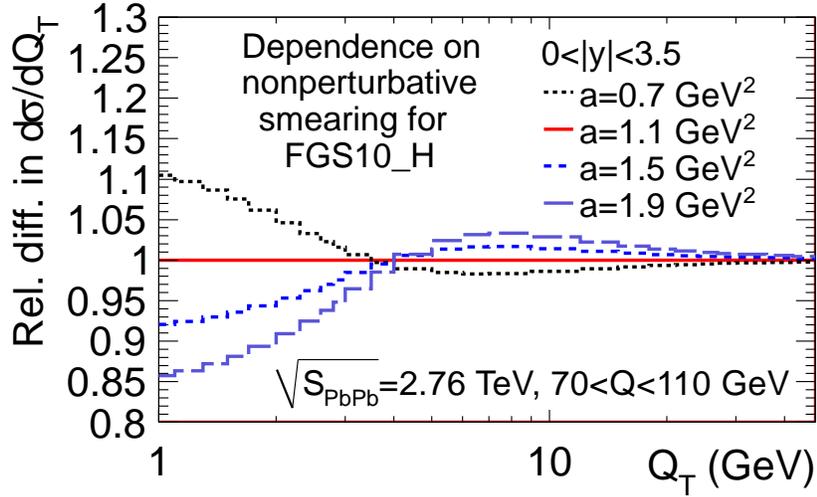}
\caption{Same as Fig.~\ref{fig:adepQ}, for $70<Q<110$ GeV.}
\label{fig:adepZ}
\end{center}
\end{figure}

Similarly, in Fig.~\ref{fig:adepZ} we show predictions 
for $d\sigma/dQ_T$ at $70 < Q < 110\mbox{ GeV}$ 
for the range $a=0.7, 1.1, 1.5,$ and $1.9\mbox{ GeV}^2$, divided by 
$d\sigma/dQ_T$ for the nominal $a=1.1\mbox{ GeV}^2$.
The range $0.7-1.5\mbox{ GeV}^2$ is of order of the
current 95\% experimental uncertainty 
in $a(Q)$ in $pp\rightarrow ZX$ \cite{Guzzi:2012jc}. 
The curve for $1.9\mbox{ GeV}^2$ is an example of 
more extreme broadening than in $pp$ collisions. Note that $Q_T$
distributions in $Z$ production are generally less sensitive 
to the nonperturbative smearing than in low-$Q$ Drell-Yan process. 

In the two figures, we observe that the strongest dependence on $a(Q)$
occurs at $Q_{T}< 4$ GeV, where the cross section varies between 80
and 150\% (85 and 130\%) in the low-$Q$ (high-$Q$) region. The
variations are much milder ($\pm 10$\% and $\pm 5$\%) at $4 < Q_T <20$ GeV
and practically vanish at $Q_T>20$ GeV. We see that, while the nonperturbative
smearing is relevant at $Q_T$ below 5 GeV, it leaves the $Q_T$
distributions intact at larger $Q_T$ and hence will not
influence sensitivity to the nuclear PDF corrections in most of the
$Q_T$ range.

\section{Conclusions\label{sec:conclusions}}

We studied production of the neutral $Z$ and $\gamma^{\ast}$ gauge bosons with their  
subsequent decay into lepton pairs in proton-lead and lead-collisions at the LHC and 
examined the role of nuclear medium modifications of nuclear PDFs. By extending 
the \textsc{ResBos} code for the Collins-Soper-Sterman $Q_T$ resummation 
to the case of nuclear parton distributions, we analyzed the 
transverse momentum $Q_T$ and rapidity $y$ dependence of the production cross section.
We examined $r_{\sigma}$, the ratio of the differential production
cross sections with and without nuclear modifications of quark and gluon PDFs in nuclei,
and found unambiguous correspondence between the predicted behavior of $r_{\sigma}$ and 
the pattern of nuclear modifications of nuclear PDFs. At central or moderate
rapidities, there may be a region with $r_{\sigma}>1$ as a consequence of antishadowing of the valence quark distributions
in nuclei. For forward and backward rapidities, we generally predict that  $r_{\sigma}< 1$ due to 
the suppression of the quark and gluon distributions in nuclei by nuclear shadowing.
For large rapidities  the ratio $r_{\sigma}$ can also discriminate between different scenarios 
(magnitudes) of nuclear corrections, while the large-$Q_T$ region probes the gluon nuclear correction.
A variety of measurements that can be carried out with $pA$ and $AA$ Drell-Yan production make it
an informative test of the QCD evolution in high-energy heavy-nuclei scattering and of 
the pattern of medium modifications of nuclear PDFs. 

The ResBos code and resummed grids for simulations 
in this study can be downloaded from http://hep.pa.msu.edu/resum/."

\section*{Acknowledgments}
M.G. and P.N. thank F. I. Olness and C.-P. Yuan for useful discussions. 
M. S. and P.N. thank organizers of 
the 2011 workshop ``High-energy QCD after the start of the LHC'' at the
Galileo Galilei Institute in Florence, Italy for financial support and
hospitality during the initial part of this work. 
This work was supported by the U.S. DOE
Early Career Research Award DE-SC0003870 and by the Lightner Sams Foundation.

%\bibliographystyle{h-elsevier3}
%\bibliography{internal}

\newpage

\end{document}